\documentclass[useAMS,usenatbib,usegraphicx]{mn2e} 

\title[Properties of Bars and Bulges in the Hubble Sequence]{Properties of Bars and Bulges in the Hubble Sequence}
\author[E. Laurikainen, H. Salo, R. Buta and J. Knapen]{E. Laurikainen$^{1}$\thanks{E-mail:
eija.laurikainen@oulu.fi}, H. Salo$^{1}$, R. Buta$^{2}$, J. H. Knapen$^{3}$\\
$^{1}$Division of Astronomy, Department of Physical Sciences, University of Oulu, FIN-90014, Finland\\
$^{2}$Department of Physics and Astronomy, University of Alabama, Box 870324, Tuscaloosa, AL 35487\\
$^{3}$Instituto de Astrof\'\i sica de Canarias, E-38200 La Laguna, Spain}
\begin{document}

\date{Accepted: Received:}


\maketitle

\label{firstpage}

\begin{abstract}

Properties of bars and bulges in the Hubble sequence are discussed, based
on the analysis of 216 disk galaxies (S0s and spirals from NIRS0S and 
OSUBGS surveys, respectively).
%
%
For that purpose we have collected together, and completed when necessary, the various analysis
we have previously made separately for early and late types.
We find strong photometric and kinematic evidence of pseudobulges
in the S0-S0/a galaxies: their bulges are on average fairly exponential,
inner disks are common (in 56$\%$), and in many of the galaxies the bulges 
are rotationally supported. This would be difficult to understand in such
gas poor galaxies as in S0s, if these pseudobulge candidates were formed by star formation 
in the disk in a a similar manner as in spirals. A more likely explanation
is that pseudobulges in the early-type galaxies are bar-related structures, 
connected to the evolution of bars, which interpretation is supported by
our Fourier analysis and structural decompositions. 
Bars in the early-type galaxies are found to have many characteristics 
of evolved systems: (1) they have flat-top/double peaked Fourier amplitude profiles,
(2) bars have typically sharp outer cut-offs, (3) the higher Fourier modes appear in the 
amplitude profiles, and (4) many bars have also ansae-type morphologies. 
We show the distributions of bar strength in different Hubble type bins using
four bar strength indicators, $Q_g$, $A_2$, $f_{bar}$ and the bar 
length, which are expected to give important clues for understanding the 
mechanism of how bars evolve.

\end{abstract}


\section{Introduction}

One central goal of extragalactic research is to understand what factors 
drive the morphology of galaxies and how important these factors are
over long term evolution. Our understanding of galaxy formation and evolution is still 
largely dominated by the early scenarios of dissipative monolithic collapse 
(Eggen, Lynden-Bell $\&$ Sandage 1962), 
and the hierarchical clustering of galaxies (Toomre $\&$ Toomre 1972; 
Toomre 1977; Navarro $\&$ Steinmetz 2000). For instance, the bulges in disk 
galaxies are assumed to have formed either in a single episode of gravitational collapse, 
or more gradually, via mergers of disk galaxies in the distant past. This
results in a classical velocity supported bulge.
Hierarchical clustering of the Universe, 
dominated by cold dark matter, seems to account remarkably well for the 
large scale structure of galaxies, but it is faced with
problems while trying to explain the structural components of individual 
galaxies. For example, galaxies are not found to have cuspy halos
(see Bosma 2003), nor is the observed amount of mass in the bulge component 
as large as predicted by these models. 
The discovery that a large majority of the bulges
in late-type spiral galaxies are actually pseudobulges formed by secular evolutionary 
processes has also been difficult to 
explain in the hierarchical picture of galaxy formation
(see the review by Kormendy $\&$ Kennicutt 2004; hereafter KK2004).

The monolithic collapse scenario of galaxy formation is also faced
with problems when compared with the recent cosmological models and observations.
It is particularly difficult to explain the so called ``downsizing'' effect  
(Cowie, Songalia $\&$ Hu 1996), which on the other hand can be understood naturally 
in the hierarchical picture of galaxy formation. There is evidence that 
the most massive galaxies were formed in the early phases of the Universe, 
having only modest star formation in the disk 
after the principal starburst event, whereas 
dimmer galaxies, in which most of the gas was not washed out in the 
main starburst event, continue actively forming stars until 
redshift zero (Cowie et al. 1996; Ueda et al. 2003; Steffan et al. 2003). 
Continued star formation in the central regions of dimmer galaxies might lead to the 
gradual formation of pseudobulges with young stellar populations. The fact that
pseudobulges have been found repeatedly in late-type spirals 
(Andreakis, Peletier $\&$ Balcells 1995; Carollo, Stiavelli 
$\&$ Mark 1998), a class dominated
by low luminosity spirals, fits to this picture. As shown by 
Laurikainen, Salo $\&$ Buta (2005), however, massive early-type
disk galaxies might also have disk-like bulges fairly frequently. Disk-like 
rotationally supported bulges have been recently detected also kinematically 
in S0 galaxies by Cappellari et al. (2005).
As it is highly improbable that these disk-like bulges were formed by star formation 
in the disk, some other mechanism is needed to explain them.
Although not yet included explicitly in cosmological models, the intrinsic
evolution of bars is among the most promising alternatives to explain 
pseudobulges in the early-type disk galaxies.

Theories of bar formation form an active field of investigation
and many interesting results have been obtained. One of the most important
discoveries has been that bars evolve due to angular momentum transfer between
bars and dark matter halos, so that bars become longer and more massive 
when evolved over time (Athanassoula 2003). 
In the early phases of evolution
a bar develops a centrally concentrated inner part and a boxy/peanut-shaped
structure, components which in dynamical models are generally interpreted as  
bulges. However, these models have not yet been fully tested with
observations. For example, it is unclear to what extent the bulges
in galaxies are classical bulges or pseudobulges, or what is the exact role of bars 
in the formation of pseudobulges? Another open question is what are the details  
of the angular momentum transfer mechanisms between bars and halos. 
This is because the dark matter halos and the bar pattern
speeds are not well known in galaxies, and also because no systematic comparison of the 
observed and model-predicted properties of bars and bulges, covering
all Hubble type bins, has been done to date. The observed properties of bars
in edge-on spirals have been compared with the dynamical models by L\"utticke,
Dettmar $\&$ Pohlen (2000) and Bureau et al. (2006), but a similar comparison has  
not been made for more face-on galaxies nor for S0s.
 
In this study the properties of bars, bulges and ovals in the Hubble sequence 
are studied and compared with various dynamical models in the literature.
In order to characterize the nature of bulges we use a photometric approach 
by applying a 2D multicomponent decomposition method (Laurikainen, 
Salo $\&$ Buta  2005), and for a subsample of galaxies stellar kinematic 
observations are also collected from the literature. The properties of 
bars and ovals in the same galaxies
are derived by Fourier techniques. The decompositions and most of the 
Fourier analysis for the individual galaxies have 
been published in a series of papers by us (Laurikainen et al.
2004, 2006a; Buta, Salo $\&$ Laurikainen 2004; Laurikainen, Salo $\&$ Buta 2004, 2005; 
Buta et al. 2005, 2006). Our aim in this paper is to combine all the analysis 
for the early- and late-type galaxies, and bring them together for a more general discussion 
in the Hubble sequence. Also, if any of the discussed parameters were 
not derived earlier for the whole sample of 216 galaxies, the analysis is
completed in this study. Such parameters are the ellipticities and the
lengths of the bars for the early-type galaxies, and the Fourier amplitude
profiles and bar morphologies for the late-type galaxies.   
Some preliminary results of this study have been reported in the IAU 
conference proceeding by Laurikainen et al. (2006). 

\section{Sample}

The sample used in this study consists of 216 galaxies, based on the Ohio 
State University
Bright Galaxy Survey (OSUBGS; Eskridge et al. 2002) for spirals, and on
the Near InfraRed S0 galaxy Survey (NIRS0S; Laurikainen, Salo $\&$ Buta 2005,
Buta et al. 2006) for S0-S0/a galaxies. Both are magnitude limited samples 
($B_T < 12.5$) avoiding high inclination galaxies ($INC < 65^\circ$). The 
NIRS0S sample has been observed only partly so far, but when combined with the OSUBGS, 
a reasonable sized sample is obtained to study 
the properties of bars and bulges in the Hubble sequence. 
Figure 1 (middle panel) shows the mean galaxy brightness in each Hubble 
type index bin.
The magnitudes are total $K_s$-band magnitudes to the
limiting surface brightness of $20 \ mag \ arcsec^{-2}$, 
corrected to Galactic extinction. The magnitudes are from Two-Micron
All Sky Survey (2MASS) \footnote{Two Micron All
Sky Survey, which is a joint project of the University of
Massachusetts and the Infrared Processing and Analysis
Center/California Institute of Technology, funded by the National
Aeronautics and Space Administration and the National Science
Foundation} and the 
extinction values are from the NASA/IPAC Extragalactic Database 
(NED) \footnote{NASA/IPAC Extragalactic Database (NED), operated by 
the Jet Propulsion Laboratory in Caltech.}. 
The mean absolute brightnesses of the galaxies in our sample are 
very similar for the  different Hubble types, except for galaxies 
of types Sc and later, which are dominated by dwarf galaxies. 
A large majority of the galaxies in our sample are bright: they are slightly 
brighter than the characteristic brightness $K_s$ = 23.1 in 
Schechter's luminosity function (Gardner et al. 1997).
The analysis discussed in this paper is based primarily on  $H$-band  
(the OSUBGS sample) and $K_s$-band (NIRS0S sample) images.

\section{Observational evidence of pseudobulges in the Hubble sequence}

The idea that some bulges might form by slow secular 
evolutionary processes after the initial rapid phase of galaxy formation was suggested 
already in 1982 by Kormendy. These so called pseudobulges have more exponential surface 
brightness profiles than the classical bulges, whose surface brightness profiles 
closely resemble those found in elliptical galaxies. The concept of a pseudobulge 
comes in two different flavors: Kormendy (1982; see also KK2004) has
suggested that they are disk-like structures formed by star formation
in the disk, whereas Athanassoula (2005) has suggested
two different types of pseudobulges, {\it disk-like} and {\it boxy/peanut}
bulges. In her characterization boxy/peanut bulges are 
thick inner components of bars seen in edge-on galaxies, which structures
are related to the characteristic orbital families 
of bars. Generally we accept the wider concept of a pseudobulge. 
However, the way how boxy/peanut structures manifest in less inclined
galaxies has not yet been studied in detail by simulations. Therefore
we adopt a term ``bar-related'' for all those pseudobulge candidates
that seem unlikely to result from star formation,  but which might be
related to the orbital structure in bars.  In the following potential
pseudobulges are identified using the photometric and kinematic
approaches, in a similar manner as suggested by KK2004 and Kormendy
$\&$ Fisher (2005). The main idea is that if the bulges were formed by
secular processes in galactic disks, they should remember their disky
origin, which is expected to be visible in their surface brightness
profiles, stellar populations, and stellar kinematics.

\subsection{Photometric approach}

In order to identify potential pseudobulges we use the following  
criteria: (1) the surface brightness 
profiles are nearly exponential (the shape parameter of the 
bulge $n < $ 2), 
(2) the galaxies have nuclear bars, rings, or ovals
in the disk inside the region of the bulge. In galaxies with
pseudobulges the bulge-to-total flux ratio ($B/T$) 
is generally small ($<$ 0.5), although this is not a 
sufficient criterion to confirm the disk-like nature of the bulge. 
In our analysis we have used a 2D multicomponent decomposition algorithm 
(Laurikainen, Salo $\&$ Buta 2005) which takes into account, not 
only the bulges and disks, but also multiple bars or ovals.
For the bulges a generalized Sersic function is used, and bars and ovals 
are typically fitted with a Ferrers function.
The decompositions for the individual galaxies discussed in this paper have been
presented previously by Laurikainen, Salo $\&$ Buta (2004, 2005)
and by Laurikainen et al. (2004, 2006). 

\subsubsection{Innermost structures of the disks}

In extensive studies, Carollo and her collaborators (Carollo, Stiavelli $\&$ Zeeuw 1997;
Carollo, Stiavelli $\&$ Mark 1998; Carollo et al. 2002) have shown 
that late-type spirals (Sb-Sbc) frequently have star forming nuclear  
rings, inner spiral arms and inner disks. In many cases the inner disks
are the only bulge-like structures in these galaxies, and some galaxies
have no bulge at all. A pseudobulge has been recently 
found also in one early-type spiral galaxy, NGC 7690, by
Kormendy et al. (2006). For some spiral galaxies there is also kinematic evidence 
confirming the disk-like nature of the bulge (F\'alcon-Barroso et al. 2003).
However, star formation related inner structures are expected to be less common in S0
galaxies, which are defficient of interstellar matter. 
In order to check the frequency of the inner structures we combined the 
data presented by Laurikainen, Salo $\&$ Buta (2005) and Laurikainen et al.
(2006) for S0-S0/a galaxies. We confirm the earlier result by Laurikainen et al. (2006) 
showing that even 56 $\%$ 
of S0-S0/a galaxies have either nuclear bars, nuclear disks or nuclear rings
inside the bulge. 

\subsubsection{$B/T$ flux ratios}
 
One of the main results of our decomposition studies is that the typical $B/T$ 
flux ratio is considerably smaller than generally assumed, particularly
for the early-type galaxies. In Figure 1 (upper panel) this is shown, for 
the first time, for all Hubble types.
For comparison, the mean $B/T$ flux-ratios derived by 
Simien $\&$ de Vaucouleurs (1986) in the $B$-band are also shown. 
The difference to our result is very large: for example, 
while for S0-S0/a galaxies Simien $\&$ de Vaucouleurs found $<B/T>_B$ = 0.57, 
we find $<B/T>_K$ = 0.25 $\pm$ 0.10. One may ask whether this is a 
wavelength effect, related to the mass-to-luminosity ($M/L$) ratio, 
or due to the different decomposition methods used?
If it is a wavelength effect we should conclude that recent star formation
(visible in the $B$-band) in the bulge, relative to that in the disk, 
should be particularly significant in S0 galaxies, which is highly improbable.
That the difference is not a wavelength effect becomes clear also 
by comparing our result with the decompositions made by 
de Souza, Gadotti $\&$ dos Anjos (2004) in the $K$-band. Using their 
measurements (their table 1) we find $<B/T>_K$ = 0.64 for S0s, which is
very similar to that obtained by Simien $\&$ de Vaucouleurs in the $B$-band.
A correction related to different
star formation time scales at different wavelengths (taken from Schulz et al. 2003)
is fairly small, changing
the $(B/T)_K$ = 0.64 to $(B/T)_B$ = 0.54 (see Laurikainen, Salo $\&$ Buta 2005 for more details).
Obviously wavelength is not capable of explaining the large
difference in $B/T$ between this study and that by Simien $\&$ 
de Vaucouleurs. A more likely explanation is that the more sophisticated multicomponent
decomposition approach used in our study is capable of accounting for 
the effects of strong bars and ovals, 
structures which in the more simple 1D (used by Simien $\&$ de Vaucouleurs) 
or 2D bulge/disk decompositions (used by de Souza, Gadotti $\&$ dos Anjos) 
are easily mixed with the bulges. 
In fact, it was shown by Laurikainen et al. (2006) that, when applied to the 
same barred galaxies, both 1D and 2D bulge/disk decompositions give fairly 
similar high $B/T$ ratios, whereas 2D bulge/disk/bar decompositions give 
significantly lower $B/T$. Laurikainen, Salo $\&$ Buta (2005) used 
also synthetic data
to demonstrate that most of the bar flux is erroneously assigned to the bulge,
if a simple bulge/disk decomposition algorithm is applied to a system with
a prominent bar. The $(B/T)_H$ flux ratios we find for spiral galaxies are also smaller
than those obtained by Simien $\&$ de Vaucouleurs, but the differences are much smaller,
presumably because bars in these galaxies are smaller and thus affect less the decompositions.  

The mean $B/T$ flux ratios we find are smaller than typically found for 
classical bulges (see KK2004).
If we assume that the mass-to-luminosity ($M/L$) ratio
is constant in these galaxies, the derived flux ratios are also approximations 
of the relative masses of bulges. In that case our result contradicts 
the earlier understanding according to which bulges and disks 
represent approximately equal amounts of mass in galaxies (Schechter $\&$ Dressler 1987; 
Benson, Frenk $\&$ Sharples 2002). Nevertheless, a constant $M/L$-ratio
might be valid for most part of the disks,
but not necessarily in the central regions of the galaxies. 
If the bulges were redder than the disk, higher $M/L$-ratios should be used 
for bulges (Bell $\&$ de Jong 2001). The conversion of flux ratios to mass 
ratios will be made in a forthcoming paper, where colour index maps 
will be studied. 

\subsubsection{Shape and scale parameters of the bulges}

Another important finding in this study is that the shape parameter 
of the bulge, $n$, is on average smaller than or near 2 for all morphological types 
(Fig. 1, lower panel). In the generalized Sersic's
function the value $n$ = 1 corresponds to an exponential disk and
$n$ = 4 corresponds to the de Vaucouleurs type surface brightness profile.
If the concept of a disk-like pseudobulge is
accepted this result seems to contradict with the view where only bulges 
in spiral galaxies later than Sb have small enough $n$-values to 
be interpreted as pseudobulges (Carollo, Stiavelli $\&$ Mark 1998; 
see also KK2004 ). However, pseudobulges in early and late-type galaxies might be different.

It has been shown both for spirals (Andreakis $\&$ Sanders 1994) and for  
elliptical galaxies (Caon, Cappaccioli $\&$ D'Onofrio 1993), that for the 
spheroidal component the generalized Sersic's function (used also in this study) gives 
a better fit than the $R^{1/4}$-law. Using a Sersic's function for the bulge 
both Andreakis, Peletier $\&$ Balcells (1995) and de Souza, Gadotti $\&$ 
dos Anjos (2004) found large $n$-values particularly for S0 galaxies
($<n>$=3.7 and 4.1, respectively), whereas Graham (2001) found $<n>$=2 
for the galaxies with the same morphological types. The $n$-values given by Graham 
are also similar to those obtained by us. How can we understand these 
disagreeing results? One of the key issues is that the sample 
by Graham did not include barred galaxies, thus  
any problems related to the degeneracy of bars and bulges were avoided.
Most probably the relative masses of bulges were overestimated
in the decompositions by Andreakis, Peletier $\&$ Balcells thus leading 
also for an overestimate of the $n$-values. 
Balcells et al. (2003) have shown the importance of high image resolution 
when deriving the $n$-values in decompositions.
They showed that the decompositions where ground-based images  
are combined with high resolution Hubble Space Telescope (HST) images,
give considerably lower
$n$-values than the decompositions where only ground-based images are used. 
The small $B/T$ flux ratios and small $n$-values for S0-S0/a galaxies
have been previously reported by Laurikainen, Salo $\&$ Buta (2005) 
and Laurikainen et al. (2006).

We confirm the earlier result by Balcells, Graham $\&$ Peletier (2004) showing that
galaxies are not scale free, the scale parameter being the relative
mass of the bulge (see Fig. 2): both the shape parameter $n$ and the effective
radius of the bulge normalized to the scale length of the disk, 
$r_{eff}/h_R$, correlate with $B/T$. A critical $B/T$ value
seems to be 0.1, below which the bulge profiles are close to 
pure exponentials having  $n$ $\sim$ 1.

The $K$-band luminosities of these galaxies provide an interesting 
additional piece of information. For the late-type spirals the $B/T$ flux ratio
increases nearly linearly towards brighter galaxies, whereas for 
the more early-type systems (S0-Sab) 
$B/T$ seems to be independent of galaxy luminosity (Fig. 2, lower panel).
If the bulges in the late-type spirals are largely  
pseudobulges formed by star formation in the disk, the found correlation can 
be understood assuming that the bulges are getting more massive when evolved over time,
which is also the generally accepted view (KK2004).
This is also consistent with
our finding that most bulges among the late-type spirals have nearly
exponential surface brightness profiles with $n$ = 1-1.5. 

On the other hand, bulges in early-type disk galaxies 
are generally believed to have non-exponential surface brightness 
profiles, interpreted as evidence of their merger origin. 
However, as a large fraction of bulges 
also in S0 galaxies are fairly exponential we should ask how were these bulges formed? 
The fact that there is no correlation between $B/T$ flux ratio 
and the galaxy luminosity is consistent with the picture that  
they have not enough gas that could accounted for the
mass of the bulge via star formation. On the other hand, if these nearly exponential bulges
were part of the bar one would expect
some differences in the $B/T$ flux ratios and the $n$-values between the barred 
and the non-barred galaxies. Figure 3 shows that such a difference seems in fact be present. 
The barred classification here is based on the de Vaucouleurs's Atlas of Galaxies
by Buta, Corwin $\&$ Odewahn (2007). If the galaxy did not appear in the
Atlas, the classification was taken from The Third Reference Catalog of
Bright Galaxies (de Vaucouleurs et al. 1991; hereafter RC3).
We also tested whether the differences in the distributions of $n$ and
$B/T$-values between the barred and nonbarred S0's are statistically
significant.  For that purpose the non-paramateric Kolmogorov-Smirnov
test was used, and all morphological types with $T \le 2$ were grouped
together. According to this test the differences between barred and nonbarred
galaxies are real: the probability that their $n$-parameters are from the
same distribution is only 0.02. The same result applies to the B/T ratios.
A similar test for later type galaxies ($T>2$) indicated no difference between
barred and nonbarred galaxies, as anticipated from Fig. 3.

As any decomposition, also our multicomponent approach has its limitations. 
A critical question is whether the uncertainties of the algorithm itself
are so large that they alone could produce the found differences
between the barred and non-barred galaxies?  
One way of evaluating this
is to look at how much massive nuclear bars or inner ovals
can affect the $B/T$ flux ratio in the decompositions. This evaluation was done 
by Laurikainen, Salo $\&$ Buta (2005) who showed that the inclusion of 
prominent inner structures to the fit do not affect the $n$-value at all,
compared to simple bulge/disk/bar decompositions,
but can decrease the mean $B/T$ flux ratio by 0.05, which is slightly smaller 
than the difference in $B/T$ between the barred and non-barred 
galaxies. 

\subsection{Kinematic approach}

Another discriminator between the pseudobulges (whether disk-like or bar-related) and the classical  
bulges is based on the kinematic approach first suggested by Illingworth (1977):
classical bulges are supported by random motions of stars 
(Illingworth 1981), whereas pseudobulges are 
supported by rotational velocities (Kormendy 1981). 
A useful discriminator is the parameter
$V_{max}/\sigma$ versus $\epsilon$ (Illingworth 1977; Kormendy 1982), 
where $V_{max}$ is the maximum rotational line of sight velocity of the bulge measured 
from the absorption lines, $\sigma$ is the mean stellar line of sight velocity dispersion
of the bulge just outside the nucleus, and $\epsilon$ is the characteristic
ellipticity inside the radius of $V_{max}$. 
Therefore $V_{max}/\sigma$ measures the importance of rotation in 
supporting the galaxy against its self-gravity.

In Table 1 we have collected the stellar kinematic observations from the 
literature for all those galaxies in our sample that have measurements 
along the major axis of the disk. We use 
those major-axis position angles and inclinations that come out from 
our ellipse fitting to deep optical or near-infrared images using the
ellipse routine in IRAF 
(PA/INC shown in the table). The bulge-dominated regions of the disks 
were evaluated, based on our structural decompositions (the radius given in the Table). 
$V_{max}$ was then 
taken to be the maximum line of sight rotational velocity inside the bulge-dominated region.
The ellipticities inside the bulge dominated region were taken from the radial $\epsilon$-profiles.
The last column in Table 1
shows the references for the kinematic data.

The data are shown in Figure 4, where the diagonal curve   
represents an isotropic oblate-spheroid 
model taken from Binney (1978). In the early diagrams by Illingworth (1981),
Kormendy (1982), and Davies et al. (1983) the elliptical galaxies 
and bulges of disk galaxies supported by 
random velocities lie below the oblate line, whereas fast rotating 
disk-like bulges are expected to appear above that line. The reliability of this
approach has been recently verified theoretically by Binney (2005).
As the apparent ellipticity and the maximum rotational velocity of the bulge depend
also on galaxy inclination, models with different inclinations were  
calculated by Binney (2005), who showed that the inclination would
shift the data points in the  
$V_{max}/\sigma$ versus $\epsilon$ diagram almost along the isotropic oblate rotator line. 
We find that the S0 galaxies in our sample appear
both above and below the line. Of these galaxies five appear above the
line: one of these galaxies, having a high flattening is a non-barred galaxy,
whereas the other four galaxies have primary bars and some of them also 
a nuclear ring. Of the S0 galaxies below the line
five are barred and two are non-barred galaxies, one galaxy has 
a clear secondary bar, and one galaxy has inner spiral arms. 

Our emphasis here is not to do any detailed interpretation of the kinematic data 
as the study for early-type galaxies by Cappellari et al. 
(2005), based on modeling of 2-dimensional kinematic observations. However, the
kinematic data for the galaxies in our sample indicates that many
of the early-type galaxies are likely to be rotationally supported. This
is also consistent with the results obtained by Cappellari et al. (2005).

\section{Bars and ovals in the Hubble sequence}

\subsection{How prominent are bars?}

Bars are generally characterized as ``strong'' when they are {\it long} 
and {\it massive} (have large Fourier amplitudes), and also when they {\it have large ellipticities} or
the {\it tangential forces induced by bars (bar torques) are large}. 
In de Vaucouleurs' family classification (de Vaucouleurs et al. 1991)
bars are divided into strong (B) and weak (AB) bars,
based mainly on visual inspection of how prominent
the bar looks in the optical image, although the morphology of the bar
also comes into the classification in the sense that weak bars 
typically have more oval-like morphologies.
As the orbital families of bars strongly depend on the underlying gravitational 
potential, one would expect a clear correlation between the 
ellipticity of a bar and the bar torque, which is indeed found to be
the case (Laurikainen, Salo $\&$ Rautiainen 2002; Buta, Laurikainen $\&$ Salo 2004).  

There are many optical (Elmegreen $\&$ Elmegreen 1985; Martin 1995; Erwin 2005)
and infrared (Regan $\&$ Elmegreen 1997; Laurikainen, Salo $\&$ Rautiainen 2002;
Laurikainen, Salo $\&$ Buta 2004) studies showing that bars in early-type 
spirals are longer than bars in late-type spirals. 
The estimation of bar length is not trivial (see Athanassoula 2005), 
but the different bar length estimates (visual inspection, maximum in the ellipticity profile, 
Fourier phase angle, or the maximum in the force profile) seem to lead
to a similar conclusion. Bars in early-type spirals are also 
found to have smaller ellipticities (Martin 1995; Shlosman, Peletier $\&$ Knapen 2000;
Whyte et al. 2002; Laurikainen et al. 2002, 2004) and smaller bar torques 
(Laurikainen, Salo $\&$ Rautiainen 2002; Buta, Laurikainen $\&$ Salo 2004) 
than bars in late-type spirals.
 
In Figure 5 we show four different bar strength indicators 
calculated for our sample of 216 galaxies:
(1) $Q_g$ (bar torque), which is the maximum of relative tangential force in the bar 
region, normalized to the underlying
mean axisymmetric force field, (2) the ellipticity of a bar using
the $f_{bar}$ index by Whyte et al. (2002), 
(3) the relative mass of a bar, as estimated from the $m$ = 2 ($A_2$) and 
$m$ = 4 ($A_4$) Fourier amplitudes in the bar region, and 
(4) the length of a bar, as estimated from the phases of the $A_2$ 
amplitudes, normalized to the radial scale length of the disk. For $Q_g$,  
$A_2$ and $A_4$, the values are collected from  Laurikainen et al. (2004, 2006),  
and Laurikainen, Salo $\&$ Buta (2005).
The values of $f_{bar}$ for the OSUBGS sample are those derived by 
Whyte et al. (2002, see also Buta, Laurikainen $\&$ Salo 2004), 
whereas  for the galaxies in the NIRS0S sample $f_{bar}$
was calculated in this study using the formula given by Whyte et al. (see Table 2): 

\vskip 0.3cm

$f_{bar} = 2 / \pi \ [\arctan (b/a)^{-1/2} - \arctan (b/a)^{+1/2}]$,

\vskip 0.3cm

\noindent where $b/a$ is the minor-to-major axis ratio of a bar.
We used $b/a$ derived from the maximum ellipticities in the bar region, 
obtained from our radial $\epsilon$-profiles, and corrected 
for the inclination of the disk (see Abraham et al. 1999). For completeness, 
the lengths of the bars for some of the galaxies in the NIRS0S sample
were also estimated in this study based on the phases of the Fourier
amplitudes, in a similar manner as estimated previously
by us for the other galaxies in our sample. 
The strength of this comparison  
is that all bar strength indicators are calculated for the same galaxies 
using the same homogeneous database. The advantage of using IR-images 
is also well known: we are not missing bars that might be obscured by dust  
in the optical region, or where morphology might be masqueraded by pockets of
star formation. 

In Figure 5 we can look at separately the tendencies for the early (shaded region)
and late-type (non-shaded region)
galaxies. Characteristic for the spiral galaxies is that bars become
{\it stronger} towards the later Hubble types when the bar torques 
are concerned, and {\it weaker} when the lengths or the 
relative masses of bars are concerned. 
These tendencies for spiral galaxies have been previously shown by Laurikainen, Salo $\&$ Buta (2004),
and for bar torques also by Buta, Laurikainen $\&$ Salo (2004). Buta et al. (2005)
also showed that in some cases spiral arms might affect the bar torques 
(in which case $Q_b$ was used). 
Figure 6 we shows that any superposition of spiral arms with bars does
not significantly affect the found tendency for the bar torques in the Hubble sequence.
For $f_{bar}$ there might be a similar increase toward later types as found for
bar torques, but it is not as clear: actually any possible variations in
the ellipticity in the Hubble sequence are within the error bars. 
This is consistent with the result obtained by
Marinova $\&$ Jogee (2006) who found that the ellipticity of a bar
is practically independent of the Hubble type.  
In any case we find that bar torque, $Q_g$, is well correlated
with $f_{bar}$ for all morphological types (Fig. 7). This correlation
exists also if spiral-corrected $Q_b$ is used, but in that case the dispersion
is somewhat larger. The apparent
inconsistency between the bar length and the relative mass of a bar in one hand, and the bar
torque on the other hand, has been discussed by Laurikainen, 
Salo $\&$ Buta (2004) for the spiral galaxies. They showed that although bars in the 
early-type spirals are longer and more massive, the bar-induced tangential forces  
are weaker, because they are diluted by the more massive 
bulges in these systems.

On the other hand, Figure 5 also shows that 
the trends we find for spiral galaxies do not extend to S0s,
which might be an important clue while
evaluating the evolutionary history of galaxies in the Hubble sequence.
Bars in the {\it early-type S0 galaxies} are clearly shorter 
(see also Erwin 2005) and less massive 
than bars in the {\it later-type S0s or in S0/a galaxies}.
For bar torques and ellipticities 
the opposite might be true, but the number of galaxies in our sample is too small
to confirm that.
In the bar torque approach the largest uncertainty is related to the
assumed vertical thickness of the bar (Laurikainen $\&$ Salo 2002).
We used a vertical thickness
based on the empirical relation between the morphological type and
the vertical thickness of the disk, as estimated by de Grijs (1998).
Nevertheless, this uncertainty (see Fig. 6, in Laurikainen et al. 2004) is 
clearly smaller than the trend we find for bar torques in the Hubble sequence. 

\subsection{Fourier amplitude profiles of bars}

\subsubsection{Early-type galaxies}

The Fourier amplitude profiles of 26 S0-S0/a galaxies in our sample 
have been previously studied by Buta et al. (2006). They showed 
that the $A_2$ amplitude
profiles of bars can be fitted by symmetric Gaussian
functions, using either single (SG), double (DG), or 
multiple (MG) Gaussian functions. In Buta et al. bars fitted by a double Gaussian
function were found to have, on average, larger bar torques than bars
fitted by a single Gaussian function. In the following we discuss 
the other bar strength indicators and the properties of bulges in the same galaxies.
As the MG-type profiles are generally very complex we concentrate only on the
SG and DG-type bars in the following.
The mean parameters of bars and bulges for these 
galaxies are collected to Table 3. We find that DG-type bars not only 
have stronger bar torques, but that they are also more elliptical, 
longer and more massive than SG-type bars.
There is also a small difference in the properties of bulges
between these two groups of galaxies
in the sense that the galaxies with DG-type bars have slightly
less massive and more exponential bulges than the
galaxies with SG-type bars. 
Both groups of galaxies have similar
$K_s$-band luminosities, which eliminates the possibility
that the found differences were due to a magnitude bias.

In order to understand better the SG and DG nature
of bars, we picked up four characteristic 
examples of early-type galaxies and inspected their properties 
in more detail (see Fig. 8). NGC 3941 is a typical example of a galaxy with an 
SG-type bar, whereas NGC 4245, NGC 2859 and NGC 1452 have
DG-type bars. For NGC 3941 the maxima in the $A_2$, $A_4$ and $A_6$
amplitude profiles appear at the same radial distance,
a behavior which is characteristic for SG-type bars.
However, in some other SG-type bars like NGC 1440, the lower modes
are shifted towards slightly smaller radial distances.
NGC 3941 has two amplitude maxima in the $A_2$-profile, the main maximum 
belonging to the primary bar, and 
the lower maximum (at r$<$5'') to an inner disk. The radial $Q_T$-profile
follows the $A_2$ amplitude profile so that the two $Q_T$-maxima
appear at the same radial distances as the two 
amplitude maxima. The surface brightness profile of this galaxy is well fitted when
including a bulge, a disk and a bar in the fit. SG-type profiles were found to  
appear both among the primary and the secondary bars, but all clearly 
identified secondary bars have SG-type amplitude profiles.

By definition in Buta et al. (2006), DG-type bars have a broad $A_2$ amplitude profile,
of which category the galaxies NGC 4245, NGC 2859 and NGC 1452 are representative examples. 
NGC 4245 has two density peaks both
in the lower ($A_2$) and in the higher Fourier modes ($A_4$ and $A_6$). 
Again, the radial $Q_T$-profile
follows the $A_2$ amplitude profile: one can imagine that
the broad maximum is actually a superposition of two partially overlapping
$Q_T$-peaks. In the structural decomposition these two components
were fitted by two oval/bar components. The 
interpretation is that the bar has two components, namely a short inner 
bar, and a longer outer bar. 
It is highly improbable that the inner component is
an oval, because the higher Fourier modes are also visible. 
In order to identify better the components both in the amplitude profile
and in the decomposition plot, in Figure 8 the peaks from the Gaussian fitting
are also shown. The parameters for calculating these peaks were taken
from Table 3 in Buta et al. (2006).

NGC 1452 has a broad double peaked $A_2$ amplitude profile, but 
in this case the higher Fourier
modes ($A_4$, $A_6$ and $A_8$) appear clearly only at r $\sim$ 35'', which is the 
more distant component of the $A_2$ double peak. Also, the
$Q_T$-profile does not completely follow the $A_2$ amplitude profile:
the peak in the $Q_T$-profile is fairly sharp and coincides 
with the radial distance where the higher Fourier modes 
are also significant. This maximum evidently corresponds to a bar,  
a component which appears also in the surface brightness profile
at a similar radial distance. 
The strongest $A_2$ maximum appears at $r$ $\sim$ 20'', where
the higher Fourier modes are weak or absent. At this 
radial distance there is only a weak shoulder in the $Q_T$-profile.
Most probably this $A_2$ maximum corresponds to a bright oval, identified 
also in the structural decomposition. This example shows that
although an oval might have a fairly high relative mass,
it does not necessarily induce strong tangential forces.
It also shows the power of the higher Fourier modes in discriminating 
ovals from thick inner components of bars.

NGC 2859 has a very broad $A_2$ amplitude profile, but in this
case the higher Fourier modes appear only at fairly small radial
distances (r $<$ 40''). The $Q_T$-profile is extremely shallow showing only
a modest peak at the radial distance where the higher Fourier
modes are present (r $\sim$ 40''). Our conclusion is that the $A_2$-profile
is broad because the galaxy has a weak bar and a prominent oval extending outside the bar,
a conclusion made also by Buta et al. (2006) for this galaxy. The small peak
in the $A_2$ amplitude profile at r $<$ 10'', visible also in the surface brightness
profile and in the $Q_T$-profile, corresponds to a secondary bar.

In conclusion, the broad or double peaked $A_2$ amplitude profiles
of bars in the early-type galaxies correspond either to  
a two-component bar with a thick inner part and a thin outer part, or a 
superposition of a bar and an oval (small or extended).
Many DG-type bars in these galaxies are two-component  
bars. We find that the DG-type bars are on average stronger
than the more simple SG-type bars using all four bar strength indicators.
Bars and ovals can be distinguished from each other by inspecting also 
the higher Fourier modes, which are significant in bars, but not in ovals
(due to their smaller ellipticities). 
Strong bars are also found to have fairly sharp outer cut-offs
in the $A_2$-profiles, in agreement with the finding by Ohta (2002). 

\subsubsection{Late-type galaxies}

We inspected the Fourier amplitude profiles also for 
all galaxies in the OSUBGS sample as well, although due to the prominence of spiral arms the 
interpretation of these profiles is not always  
straightforward. For this reason, any quantitative comparison
of SG/DG type bars in different Hubble type bins
was not possible. However, similar density profiles as found 
for the early-type galaxies were found also for the spirals. 
Examples of characteristic SG and DG-type profiles of bars in spiral 
galaxies are collected in Figure 9. As in Fig. 8, the 
radial $Q_T$-profiles and Fourier amplitude profiles
are shown.

'The bar in
NGC~4321 manifests itself with a rather weak outer part, affected in
morphology by the spiral arms, and an aligned inner part which is
well-defined in the near-IR (Knapen et al. 1995). This
secondary bar, with a length of some 10 arcsec, has a clear
SG-type morphology.
The morphological type of this galaxy is SAB(s)bc, but the 
amplitude profile of the bar is similar as in 
any other morphological type: the density
maximum is sharp and in addition to the $A_2$ component the 
higher Fourier modes, $A_4$ and $A_6$
(even $A_8$ is present), are also prominent. All
modes appear nearly at the same radial distance, which is also
the location of the $Q_T$-maximum. 

The five remaining galaxies in Figure 9 are candidates of DG-type 
bar profiles, similar to those discussed 
among the early-type galaxies above. For example, NGC 3583 clearly has 
a two-component maximum at r $<$ 40'', 
both in the lower and in the higher Fourier modes, indicating
that the bar has both an inner and an outer component. 
The $Q_T$-peak is also very broad.
However the phase of the $A_2$ component is maintained
nearly constant only to $r$ $\sim$ 25'', which suggests that the outer part of the bar has
some spiral-like characteristics, as is visible also in the 
direct image of this galaxy. 
The galaxies NGC 4394 and NGC 7479 are more clear examples of 
two-component bars. Bars in these galaxies have
qualitatively similar double-peaked amplitude profiles with prominent higher 
Fourier modes. The bar in NGC 4593 is qualitatively 
similar to those in NGC 4394 and NGC 7479. The large $A_2$ maximum at $r$ $\sim$ 70'' is caused 
by the prominent spiral arms starting at the two ends of the bar.

Ohta, Hawabe $\&$ Wakamasu (1990) were the first to discover 
that the higher Fourier modes of bars are 
characteristic for early-type galaxies, like S0s. They also argued that
these modes are absent for the bars in spiral galaxies. 
However, our analysis indicates that the higher Fourier modes 
are characteristic for {\it all} strong bars,
independent of the morphological type. For example, the 
amplitudes of the higher modes are extremely strong 
in the Sc-type spiral galaxy NGC 7479, 
having $Q_g$=0.7, and weaker for NGC 6221 
($Q_g$ =0.44). 
According to Ohta (1996) the early-type galaxies also have sharp cut-offs
in the amplitude profiles at the ends of the bar, but such
cut-offs were argued to be missing in bars of late-type galaxies.
Due to the strong spiral arms, the shapes of the amplitude profiles in 
the spiral galaxies are difficult to evaluate. However, NGC 7479 
is an example showing that strong bars 
can have fairly sharp outer cut-offs in their 
$A_2$-profiles also in the late-type spirals.
Most probably, this property is also related to the 
strength of the bar rather than to the morphological type.
As an example of an early-type bar Ohta used NGC 4643, a galaxy which
is also in our sample, and shows a fairly strong bar ($Q_g$=0.3). 

\section{Discussion}

\subsection{Nature of bulges in the Hubble sequence}

In the hierarchical clustering model of the Universe dominated by cold
dark matter (Toomre $\&$ Toomre 1972; White $\&$ Rees 1978; Steinmetz $\&$ Navarro 2002) 
the dominant mechanism for producing bulges is by mergers of disk galaxies.
Mergers of equal mass galaxies also yield remnants with properties 
similar to those found in the elliptical galaxies (Barnes 1988; Hernquist
1993; Lima-Neto $\&$ Combes 2005; Balcells $\&$ Gonzalez 1998).
However, a problem with these models is that the  
properties of bulges in spiral galaxies and, as discussed in this study, even  
in early-type disk galaxies, do not resemble those of the elliptical galaxies. 
Nevertheless, it is not clear how strict the criterion for the merger
origin is: there is recent evidence based on the cosmological 
dynamical models by Springer $\&$ Hernquist (2005) showing that if
sufficient gas remains following a major merger, cooling 
can quickly reform the disk. This yields remnants that are
closer to spiral galaxies, both structurally and kinematically.
If this is correct, bulges 
might still be remnants of hierarchical clustering, particularly 
in the non-barred early-type galaxies. However,
detailed comparison between observations and  
model predictions are not yet possible. 

The fact that the bulges in the spiral galaxies are nearly exponential
is consistent with the picture according to which the bulges are largely
part of the disk formed by star formation in the disk. The $B/T$-ratios are
also small and
$B/T$ flux ratio increases with  galaxy luminosity. 
This can be understood if gas from the outer disk
is accreted to the inner parts during the galaxy evolution. Galaxies 
with higher luminosities have more gas in the disk that can 
accumulate into starbursting rings and 
eventually lead to increased bulge masses. Support for this scenario comes 
from the recent Spitzer Space Telescope
observations by Fisher (2006) who showed that galaxies which are 
structurally identified as having pseudobulges, also have 
higher central star formation rates than those
having classical bulges. 
Bars may help in transferring the gas towards the central regions of the galaxies,
but they are not a necessary requirement for the gas inflow (Sakamoto et al. 1999; Sheth et al. 2005).

The predominance of old stellar populations and
the de Vaucouleurs type surface brightness profile are generally used
to claim that bulges in S0-S0/a galaxies are merger-built structures (Schweizer 2005).
Also, they do not seem to have sufficient gas for 
producing star formation at a level that could account for the masses of
typical pseudobulges (see KK2004). However, as discussed in this study, 
bulges in the early-type galaxies have many signatures of pseudobulges.
For example, they have fairly exponential bulges, 56$\%$ of them have inner 
structures like nuclear rings or nuclear bars, and 
some of them also have kinematic evidence of rotationally supported bulges.
There is also a large kinematic study of E/S0 galaxies by Emsellen et al. (2005, and
references there) using high resolution integral field spectroscopy for deriving
the kinematic parameters.
They found that the bulges in most S0s galaxies 
are fast rotating systems having also large anisotropies of the bulge.
These kinematic properties were suggested to indicate either
secular evolution, or heating of the disk due to minor mergers. 
Our interpretation in this study is that pseudobulges in the early-type disk
galaxies are largely bar-related, connected to the evolution of bars. This could also
naturally explain why the bulges are non-classical even in such gas poor galaxies
as S0s.

\subsection{Angular momentum transfer and the evolution of bars in the Hubble sequence}
 
In modern dynamical models, bars are expected to play an important 
role in galaxy evolution (Athanassaoula 2003). The main idea is that bars evolve due to angular momentum
transfer between the bar and the halo, which occurs
near the resonances: in particular disk material at the Inner Lindblad Resonance (ILR)
will lose angular momentum, while halo material near corotation (CR) and
near the outer Lindblad Resonance (OLR) will absorb it. How strong this
angular momentum exchange is depends critically on the mass of the
halo and its central concentration, the amount of mass in the resonances,
and how dynamically cool or hot are the disk and the halo. 
Athanassoula (2003) reports three different ways of how bars can lose angular 
momentum: (1) by trapping particles outside the bar into elongated
orbits of the bar, a process in which angular momentum is lost from the
inner parts of the disk, while the bar becomes {\it longer}, (2) part or
all of the orbits trapped in the bar become more elongated and the bar 
becomes {\it thinner}, and (3) bars lose rotational energy leading to 
a {\it slow bar}. These three processes are expected to be linked, 
so that due to strong angular momentum transfer bars simultaneously
become longer, thinner and more slowly rotating. Athanassoula's models (2003)
have also shown that while bars become longer, their relative masses
increase. These trends have been verified later by other  
self-consistent 3D simulation studies like those made by
Martinez-Valpuesta, Shlosman $\&$ Heller (2006). The first indication 
that bars grow when they evolved over time comes
already from dynamical models by Sellwood (1980).

Our observations show that bars become longer and more massive from the 
late-type spirals towards the early-type spirals. 
These observations are 
consistent with the models by Athanassoula (2003) if either the dark matter halos 
or the bulges in the early-type galaxies are more massive or 
more centrally concentrated than the halos and bulges in late-type galaxies. 
Our observations can be understood by the same
models also if the bars in the early-type galaxies are very old structures,
in which case even smaller halos might be sufficient to cause the angular momentum transfer.
In that case we might also be witnessing slow evolution of bars in galaxies 
that gradually lose their gas and 
change their morphological type in the Hubble sequence. A possible
candidate of such evolution is the non-barred galaxy NGC 1411, which is 
classified as an S0 galaxy, but has a  $B/T$ flux ratio which is as 
small as typically found in Sc-type spirals (see Laurikainen et al. 2006). 

On the other hand, and 
at odds with the model predictions, our analysis also shows that bars 
do not have larger bar torques
nore become more elliptical, towards the early-type spirals. 
This behavior of the bar torque ($Q_g$) and the ellipticity of the bar 
($f_{bar}$) however, should not necessarily 
be taken as a counterargument to the evolutionary models by Athanassoula. This is because  
$Q_g$ is diluted by the underlying axisymmetric component, generally
the bulge, which is more massive in the early-type galaxies, and 
which is generally not taken into account in the simulations.
It would be interesting if the simulations
could confirm whether the opposite trends found in the Hubble sequence among the different 
bar strength indicators can be completely explained by the dilution effect
due to more massive bulges in the early-type galaxies as suggested by 
Laurikainen, Salo $\&$ Buta (2004), or whether this result is also related 
to the different mechanisms of how bars lose their angular momentum.

An additional verification for the hypothesis that bars evolve because
they lose angular momentum would be to show that long, evolved bars
are more slowly rotating systems than shorter and less evolved bars.
However, this has been difficult to prove, mainly because of the
difficulties to measure the bar pattern speed, $\Omega_p$. So far,
$\Omega_p$ has been measured directly using the Tremaine-Weinberg
method for some S0 galaxies (Merrifield $\&$ Kuijken 1995: Gerssen,
Kuijken $\&$ Merrifield 2003; Aguerri, Debattista $\&$ Corsini 2003;
Corsini, Debattista $\&$ Aguerri 2003; Rand $\&$ Wallin 2004;
Debattista $\&$ Williams 2004), and only for two spiral galaxies
(Gerssen, Kuijken $\&$ Merrifiled 2003). These direct measurements of
$\Omega_p$ have repeatedly pointed to fast bars in the early-type
galaxies, contrary to what one would expect for evolved bars.  The
largest collection of $\Omega_p$ measurements for spiral galaxies
comes from the 2D sticky particle simulation (Salo et al. 1999) models
for 38 spiral galaxies in the OSUBGS sample by Rautiainen, Salo $\&$
Laurikainen (2005) and Salo et al. (2006), who showed that the bars in
late-type spirals are actually slower rotators than those in
early-type spirals. In principle, this could be related to the dark
matter halos, because according to Persic, Salucci $\&$ Stel (1996)
the relative halo mass depends on galaxy luminosity so that galaxies
with lower luminosities also have more massive dark matter
haloes. However, in the OSUBGS sample the luminosities of Sb and Sbc
galaxies, for which lower $\Omega_p$-values were found, have on
average the same luminosities as the early-type spirals. Therefore,
mass of the dark matter halo alone is not sufficient to explain the
slowdown rate of the bar in the late-type spirals.  More measurements
of $\Omega_p$ and of the rotation curves are evidently needed to
clarify this issue.

A completely different view of the evolution of bars in spiral galaxies 
has been presented by Bournaud $\&$ Combes (2002), who suggest that 
in the presence of continuous accretion of external gas bars become
fairly short-lived systems, so that bars are recurrently formed
and destroyed.
In principle this is possible, but our observations do not shed any 
new light on this issue. The recent simulations by Debattista et al. (2006)
suggest that bars are actually fairly robust systems so that high gas masses 
are required to destroy the bars. 
Somewhat smaller gas masses to destroy bars are suggested in the
simulation models by Hozumi $\&$ Hernquist (2005), Bounaud, Combes $\&$ Semelin (2005),
and Athassoula, Lambert $\&$ Dehnen (2005). 

Bar strength measurements alone are not sufficient to distinguish 
whether bars are strong mainly because they have massive dark matter 
halos, or because the strong bars are very old, formed in the epoch 
when they still had a large amount of gas. Whatever the case, something
is different in galaxies with S0/a type morphologies. For very early types
the bars start to decrease in length and lose their mass, probably accompanied 
by smaller bulge components. An intriguing possibility is that the
internal evolution of bars plays an important role in producing these characteristics.  

\subsection{Internal evolution of bars}

The evolution of stellar bars is affected by dynamical instabilities
leading to long-term changes in their morphologies. A well known 
is the so called buckling instability (Combes et al. 1990; Pfenniger $\&$ Friedli 1991;
Raha et al. 1991; Berenzen et al. 1998; Athanassoula $\&$ Misioritis 2002; 
Athanassoula 2002, 2003, 2005a,b; O'Neilss $\&$ Dubinski 2003; 
Debattista et al. 2004; Martinez-Valpuesta, Shlosman $\&$ Heller 2004; 
Debattista et al. 2006; Martinez-Valpuesta, Shlosman $\&$ Heller 2006),
where the orbital families of bars are changed in such a manner that 
leads to a vertical thickening of the bar, the so called boxy/peanut 
structure. In recent studies 
(Martinez-Valpuesta, Shlosman $\&$ Heller 2006; Athanassoula 2006) 
multiple buckling effects have also been discussed. Buckling is expected
to be a natural part of the evolution of bars, being particularly important
in strong bars. The presence of boxy/peanut bulges for edge-on 
galaxies has previously been shown both kinematically (Kuijken $\&$ Merrifield 1995; 
Chung $\&$ Bureau 2004) and using surface photometry (L\"utticke, Dettmar
$\&$ Pohlen 2000; Bureau et al. 2006). L\"utticke, Dettmar $\&$ Pohlen found that
54$\%$ of the bulges of all morphological types among edge-on galaxies 
have boxy/peanut bulges, which implies that they should be common
also among the less inclined galaxies.

When a bar forms in numerical simulations it is thin, but soon develops a 
vertically thick inner part and a vertically thick more extended middle component 
of the bar (Athanassoula 2003, 2005). In the surface brightness profiles
both vertically thick components can be interpreted as
pseudobulges. In the simulation models the surface brightness
profile of the innermost component of the bar is nearly exponential, whereas
for the middle component it depends on
the viewing angle: when viewed edge-on the profile takes
a Freeman type II profile shape. Also the length of the
peanut structure depends on the viewing angle so that it is
longest in the edge-on galaxies (Athanassoula $\&$ Beaton 2006). The strength of the peanut structure 
depends on bar strength so that it is strongest in strong bars
(Athanassoula $\&$ Misioritis 2002; Bureau $\&$ Athanassoula 2005).
Strong bars in simulation models are also found to have frequently ansae-type
morphologies (Athanassoula $\&$ Misioritis 2002; Athanassoula, Lambert $\&$
Dehnen 2005; Athanassoula $\&$ Beaton 2006; Martinez-Valpuesta, 
Shlosman $\&$ Heller 2006).

However, a comparison of the observations with the simulation models
is difficult because different parts of the bars are seen in face-on
and in edge-on views. This has been demonstrated for example by
Athanassoula $\&$ Beaton (2006), who showed that although the
boxy/peanut structure is visible at an inclination of 77$^0$, the
length of this structure is shorter at this inclination than in the
edge-on view. Also, in face-on view the boxy/peanut structure is
barely visible.  Based on these difficulties we cannot argue that the
bulges we see in the early-type galaxies are boxy/peanut structures
produced by buckling effects.  However, we have shown strong evidence
that bars in the early-type galaxies are evolved systems, which is
also the case with the boxy/peanut shaped bar/bulges in the simulation
models.  Namely, we find that 90$\%$ of early-type galaxies have
either flat or intermediate-type Fourier amplitude profiles, and
40$\%$ have ansae bar morphologies. In comparison, the amplitude
profiles in spiral galaxies are mostly exponential and only 12$\%$
show ansae-type morphologies (14 with ansae among 115 OSUBGS galaxies).
\footnote{a similar
result for the early-type galaxies has been obtained also by Martinez-Valpuesta $\&$ Knapen 2007,
private communication}, 
Also, by comparing the Fourier amplitude profiles and the multicomponent decompositions 
for the same galaxies we identify inner structures that cannot be explained
by the classical bulges or ovals.
Buta et al. (2006) showed that the observed flat-top amplitude 
profiles are at least consistent with models with large dark matter halos, 
but other factors as centrally concentrated halo mass profiles or enough 
mass in the resonances might also lead to similar evolved amplitude profiles.

Therefore, both the morphological analysis of bars and bar strengths in S0-S0/a 
galaxies hint to evolved bars. However, what still needs to be explained 
in this picture is why bars in the S0-S0/a galaxies are
repeatedly found to be fast? And also, why bars, particularly in
the early-type S0s, are less prominent than bars in the later type S0s or in S0/a galaxies?
None of this is expected in models where bars evolve due to a significant transfer
of the angular momentum from the bar to the halo.
It has been shown by
Martinez-Valpuesta, Shlosman $\&$ Heller (2006) that the connection between 
the different properties of bars might actually be a complicated process,
so that during the buckling process $\Omega_p$ sharply increases,
followed by a sudden decrease in bar length. However, 
even in this model the bar should both grow in length and slow down
in the long run, so that after the 
second buckling the bar should be more prominent and more slowly rotating 
than after the first buckling. 

There are other models, however, like those by Athanassoula, Lambert and
Dehnen (2005), that can account better for the observed properties
of bars in the S0 galaxies. They showed that if central mass concentrations
like realistic secondary bars or nuclear disks are used in the models, 
bars become shorter, less massive and faster than in models 
without any centrally concentrated mass components. Nuclear bars
and inner disks appear in more than 50$\%$ of the studied S0-S0/a galaxies
in our sample, implying that the models by Athanassoula, Lambert $\&$ Dehnen are
a promising explanation for the type of bars we find in the very early-type galaxies.
In the context of evolution 
it would be natural to expect that the early-type
S0 galaxies are older than the late-type S0s or S0/a galaxies,
which implies that their bars have had more time to evolve:
in this case to become less prominent.   
Ultimately, the efficiency of nuclear bars and disks in redistributing 
matter in the disk is related to the problem of cuspy halos: if the
halos are cuspy as assumed in present cosmological models, the central
parts of the galaxies are halo dominated, thus reducing the effects
discussed by Athanassoula, Lambert $\&$ Dehnen (2005). But if the halos
have constant density cores then their models might efficiently produce the type of
bars found in S0s.

Although it seems that bars in early-type galaxies have many characteristics 
of evolved systems, the details of the mechanism of the
angular momentum transfer between the bar and the dark matter halo
remains unsolved. Even if bars evolve due to the angular momentum
transfer between the bar and the halo 
the time scale of the evolution can be either rapid or more slow, 
which is expected to have imcplications to the ages of bars and 
the masses of dark matter haloes. 
The masses of the halos in the early-type galaxies
are poorly known, but the rotation curve observations 
by Mathieu, Merrifield $\&$ Kuijken (2002, see also Romanowsky 2006) seem to indicate
that the halo masses in S0 galaxies might actually be quite small.
Bulges are expected to affect the evolution of bars in a similar manner
as the dark matter halos (Athanassoula 2003), but this has not yet been tested
using realistic $B/T$-ratios for galaxies in the different Hubble type bins. 

\section{Conclusions}

Properties of bars and bulges in the Hubble sequence are discussed,
based on the analysis of 216 disk galaxies, selected
from the OSUBGS and NIRS0S samples. Our main emphasis has been to
combine the various properties of bars and bulges, presented
for the individual galaxies in a series of papers by us, and
to discuss the implications of these measurements in the Hubble sequence. 
The properties of bars were derived mainly by Fourier techniques and the properties of
bulges by applying a multicomponent decomposition code. The
analysis results were compared with various dynamical models in the literature.
We find strong photometric and kinematic evidence of pseudobulges 
in the S0-S0/a galaxies. However, most probably pseudobulges
in the early and late-type galaxies have a different origin, 
which would also make more understandable why pseudobulges are
frequently found in such gas poor galaxies as in S0s.

Our main results are the following:   

(1) We show both photometric and kinematic evidence that the bulges 
in many S0-S0/a galaxies have characteristics of pseudobulges: they
have on average small shape parameters and 56$\%$ of them 
have inner components as nuclear bars or nuclear rings, 
confirming the earlier results by Laurikainen et al. (2006). 
Kinematic data collected from the literature is in many cases
consistant with rotationally supported bulges.

(2) We find evidence showing that pseudobulges in the late-type 
galaxies were formed by star formation in the disk, while in the early-type
galaxies (S0-S0/a) they are mainly bar-related structures.
This interpretation is based on the decompositions, which show that
these components have nearly exponential surface brightness profiles, and
on Fourier analysis showing flat-top/double-peaked amplitude profiles,
where the higher Fourier modes are also important.

(3) We find that bars with flat-top/double peaked amplitude profiles are stronger
that the more simple bars, using all bar strength indicators: bar torque ($Q_g$),
the ellipticity of a bar ($f_{bar}$), the relative mass a bar ($A_2$) and the length of a bar.
Galaxies with flat-top/double-peaked amplitude profiles have also slightly more
exponential bulges and lower $B/T$ flux ratios than the more simple bars. 

(4) Strong bars are found to have the following characteristics:
1) flat-top/double peaked amplitude profiles
2) sharp outer cut-offs, 3) a presence of higher Fourier modes, and 4) in many
cases ansae-type bar morphologies. 
We found that 40$\%$ of S0-S0/a galaxies have ansae, in contrast to 13$\%$ in spiral galaxies.
Flat-top/double or intermediate-type amplitude profiles are found in 92$\%$ of 
S0-S0/a galaxies, whereas amplitude profiles of bars in spiral galaxies are largely exponential. 

(5) The distributions of bar strength in different Hubble type bins are shown using
four different bar strength indicators, $Q_g$, $A_2$, $f_{bar}$ and bar length, which
is expected to give important clues for understanding the mechanism of how bars lose
their angular momentum and thus evolve over time. For example, 
it needs to be understood why bars in the early-type S0 galaxies are less
prominent than bars in the later type S0s or in S0/a galaxies.

(7) Even 70 $\%$ of S0-S0/a galaxies are found to have ovals or lenses,
confirming the earlier result by Laurikainen et al. (2006).
According to dynamical models by Athanassoula (2003)
weak ovals may form in hot halos and more extended
ovals in hot disks. We find both weak and expended ovals and
thin classical bars in the same galaxies, which needs to be
explained by the theoretical models.

\section*{Acknowledgments}
We wish to thank Lia Athanassoula and Inma Martinez-Valpuesta for valuable 
comments to this manuscript. We also thank Anna M\"akinen for collecting the kinematic data
from the literature, as a summer work in the Astronomy Division, Univ. of Oulu. The support
from the Academy of Finalnd is acknowledged.

\newpage
\begin{table*}
   \caption[]{Stellar kinematical properties of bulges. }
\begin{tabular}{cccccccc}
\hline
\noalign{\smallskip} 

 Gal.       & T   &  INC/PA   & $V_{max}$  & $\sigma$  & $\epsilon$  & r        & ref.               \\  
            &     &  [degrees]& [$km /sec$] &[$km /sec$] &            & [arcsec]       &                               \\
\hline
\noalign{\smallskip}

            &    &            &       &           &        &       &                             \\ 
NGC 210     & 3  & 49.2/162.7 &  83.3 &120        & 0.1    & 10''  &Pizzella et al. 2004       \\
NGC 936     &-1  & 42.4/123   &  80.2 & 175       & 0.15   & 5''    & Kormendy (1983)         \\ %
NGC 1400    &-3  & 23.8/37    & 90.0  &266        & 0.1    & 5''    & Bertin et al. 1994      \\ %
NGC 1553    &-2  & 41.5/150   & 99.4  &162        & 0.4    & 10''   & Kormendy 1984            \\ %
NGC 1574    &-3  & 16/31      & 39.9  & 180       & 0.1    & 6''   & Jarvis et al. (1988)    \\ %
NGC 2681    & 0  & 24.2/102(2)& 20.1  & 111       & 0.2    &5''     & McElroy 2004; Moiseev et al.2004   \\ %
NGC 2775    & 2  & 36.8/166.5 & 125.2 &175        & 0.11   & $<$20'' & Corsini et al. 1999       \\
                 &   &        & 100.0 &140        & 0.11   & $<$20'  &Shapiro et al. 2003        \\
NGC 2855    & 0  & 33.1/107   & 120.0 & 190       & 0.17   & 10''   & Corsini et al. 2002       \\
NGC 2983    &-1  & 54.8/91    & 54.7  &160        &0.2     &2''     & Bettoni et al.1988        \\ %
NGC 3169    & 1  & 39.1/58.0  & 112.0 & 171       & 0.36   & 10''  & Heraudeau $\&$ Simien 1998       \\
NGC 3626    &-1  & 48.0/159   & 150.0 &142        & 0.4    &5''     &Haynes et al.2000         \\
NGC 3706    &-4  & 50.65/76   & 139.9 &281        &0.38    &10''    &Carollo et al.1994; 1993   \\ %
NGC 3810    & 5  & 47.2/23.1  & 50.0  & 73        & 0.27   & 15'' & Heraudeau et al. 1999       \\    
NGC 3941    & -2 & 49.4/7     & 60.0  & 131/150   &0.2     &5''     &Fisher 1997;Denicolo2005   \\ %
NGC 4138    &1   & 53.3/148.2 & 100.0 & 161       & 0.21   & 5''   &Jore et al. 1996           \\
NGC 4340    &-1  & 56.2/99    & 48.2  &115        &0.07    &5''     &Simien,Prugniel,1997        \\ %
NGC 4450    &2   & 43.9/2.8   & 45.0  & 126       & 0.25   & 10'' &  Fillmore et al. 1986       \\
NGC 4579    &2   & 38.5/95.3  & 55.0  & 174       & 0.22   & 5-10'' & Palacios et al. 1997, Heraudeau $\&$ Simien 1998  \\
NGC 4596    & 0  & 44.3/116   & 45.0  &149        &0.15    &5''     &Bettoni $\&$ Gallaher, 1997     \\ %
NGC 4643    & 0  & 33.3/50    & 79.6  &167        &0.1     &5''     & Magrelli et al.1992   \\ %
NGC 5005    &3   & 63.6/67.9  &110.0  & 213       & 0.25   & 5''  & Batcheldor et al. 2005     \\
NGC 7727    & 1  & 26.9/159.8 & 35.0  & 181       & 0.27   & 8''  & Simien $\&$ Prugniel 1997   \\
ESO 208-G21 &-4  & 43.9/109   & 110.0 & 150       & 0.52   & 5''    & Carollo et al. (1993)   \\ %

\noalign{\smallskip}
\hline
\end{tabular}

\end{table*}

\newpage

\begin{table*}
   \caption[]{The ellipticies of the barred NIRS0s galaxies, using the
parameter $f_{bar}$ as defined by Whyte et al. (2002).}
\begin{tabular}{cc}
\hline
\noalign{\smallskip}

NGC  & $f_{bar}$       \\
     &        \\  
718   &  0.124  \\  
936   &  0.273  \\  
1022  &  0.281  \\  
1079 &   0.262  \\  
1317 &   0.107  \\  
1326 &   0.220  \\  
1350 &   0.234  \\  
1387 &   0.117  \\  
1415  &  0.239  \\  
1440  &  0.173  \\  
1452  &  0.297  \\  
1512 &   0.306  \\  
1533 &   0.167  \\  
1574 &   0.120  \\  
2273  &  0.249  \\  
2681  &  0.098  \\  
2781  &  0.070  \\  
2859  &  0.235  \\  
2983  &  0.262  \\  
3081 &   0.256  \\  
3358  &  0.132  \\  
3626  &  0.142  \\  
3941  &  0.077  \\  
4245  &  0.164  \\  
4340  &  0.283  \\  
4596  &  0.241  \\  
4608  &  0.243  \\  
4643  &  0.255  \\  

\noalign{\smallskip}
\hline
\end{tabular}

\end{table*}

\newpage

\begin{table*}
   \caption[]{Comparison of galaxies having single and double gaussian bars.
The errors are mean errors.}
\begin{tabular}{ccc}
\hline
\noalign{\smallskip}

                      &    SG            & DG             \\         
                      &                  &                 \\                     
        $<M_K>$         & 23.4 $\pm$ 0.2   & 23.4 $\pm$ 0.3     \\   
        $<Q_g>$         & 0.09 $\pm$ 0.01  & 0.23 $\pm$ 0.03    \\   
        $<f_{bar}>$      & 0.17 $\pm$ 0.09  & 0.22 $\pm$ 0.09     \\             
        $<A_2>$         & 0.39 $\pm$ 0.03  & 0.61 $\pm$ 0.05    \\   
        $<barlen/h_R>$  & 1.14 $\pm$ 0.11  & 1.56 $\pm$ 0.12    \\   
        $<B/T>$         & 0.27 $\pm$ 0.04  & 0.17 $\pm$ 0.02     \\  
        $<n>$           &  1.9 $\pm$ 0.2   & 1.4  $\pm$ 0.2     \\   

\noalign{\smallskip}
\hline
\end{tabular}

\end{table*}
\clearpage

\begin{figure}
\includegraphics[]{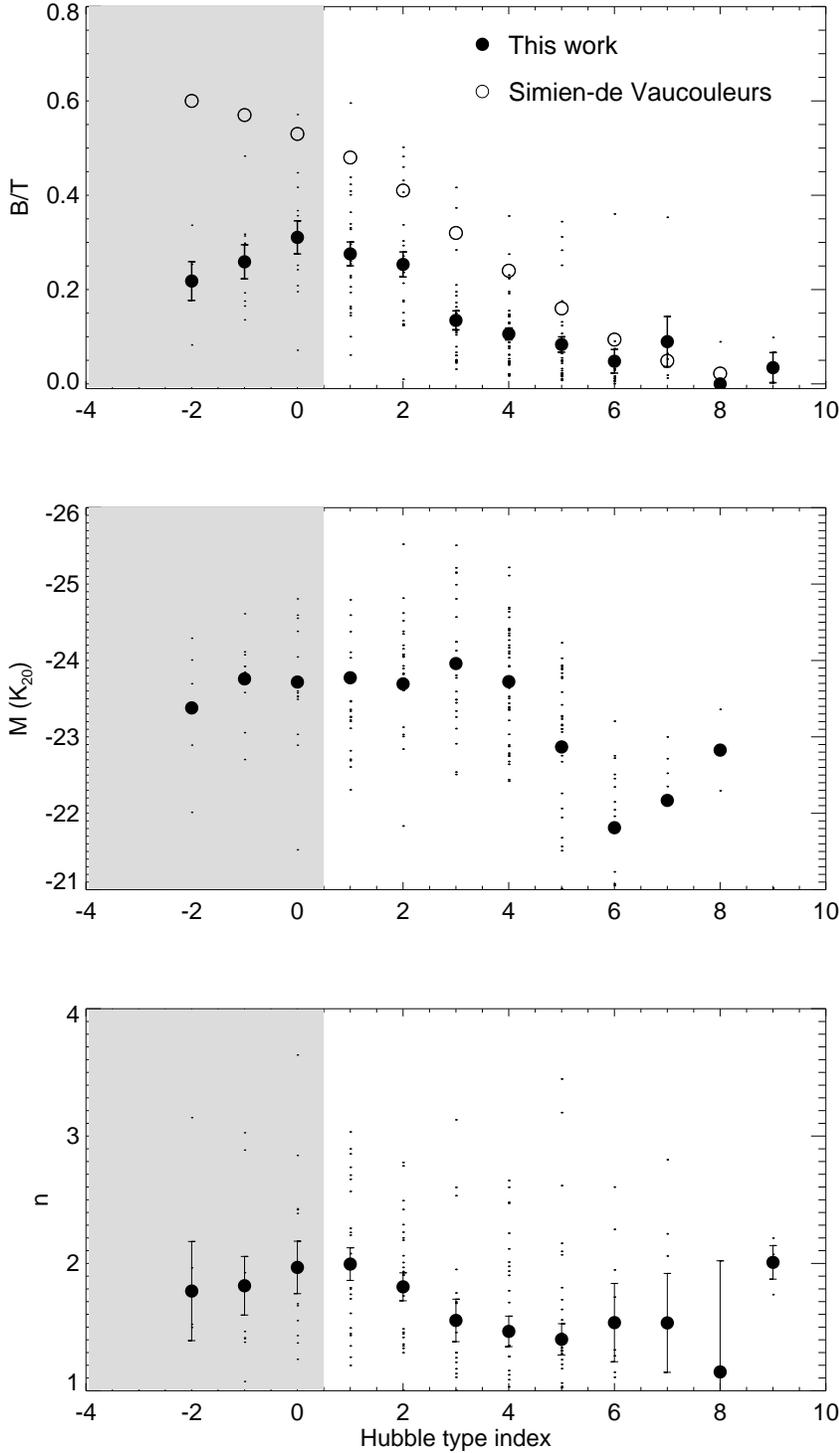}
\caption{In this figure are shown the bulge-to-total flux ratio, $B/T$ (upper panel),
the absolute $K$-band galaxy luminosity using the magnitudes to the surface brightness
of 20 $mag / arcsec^{2}$ taken from the NED and corrected for Galactic extinction (middle panel), 
and the shape parameter of the bulge, $n$ corresponding to Sersic's function (lower panel), 
as a function of the Hubble type index. These
parameters are calculated for our sample of 216 galaxies. The large symbols indicate mean values
and the error bars are standard deviations of the mean. For comparison, in the upper panel 
the measurements by Simien $\&$ de Vaucouleurs, obtained in the $B$-band (1986) are also shown.
Our parameters of the bulges are derived from $K_s$ and $H$-band images.} 
\label{sample-figure}
\end{figure}

\clearpage

\begin{figure}
\includegraphics[]{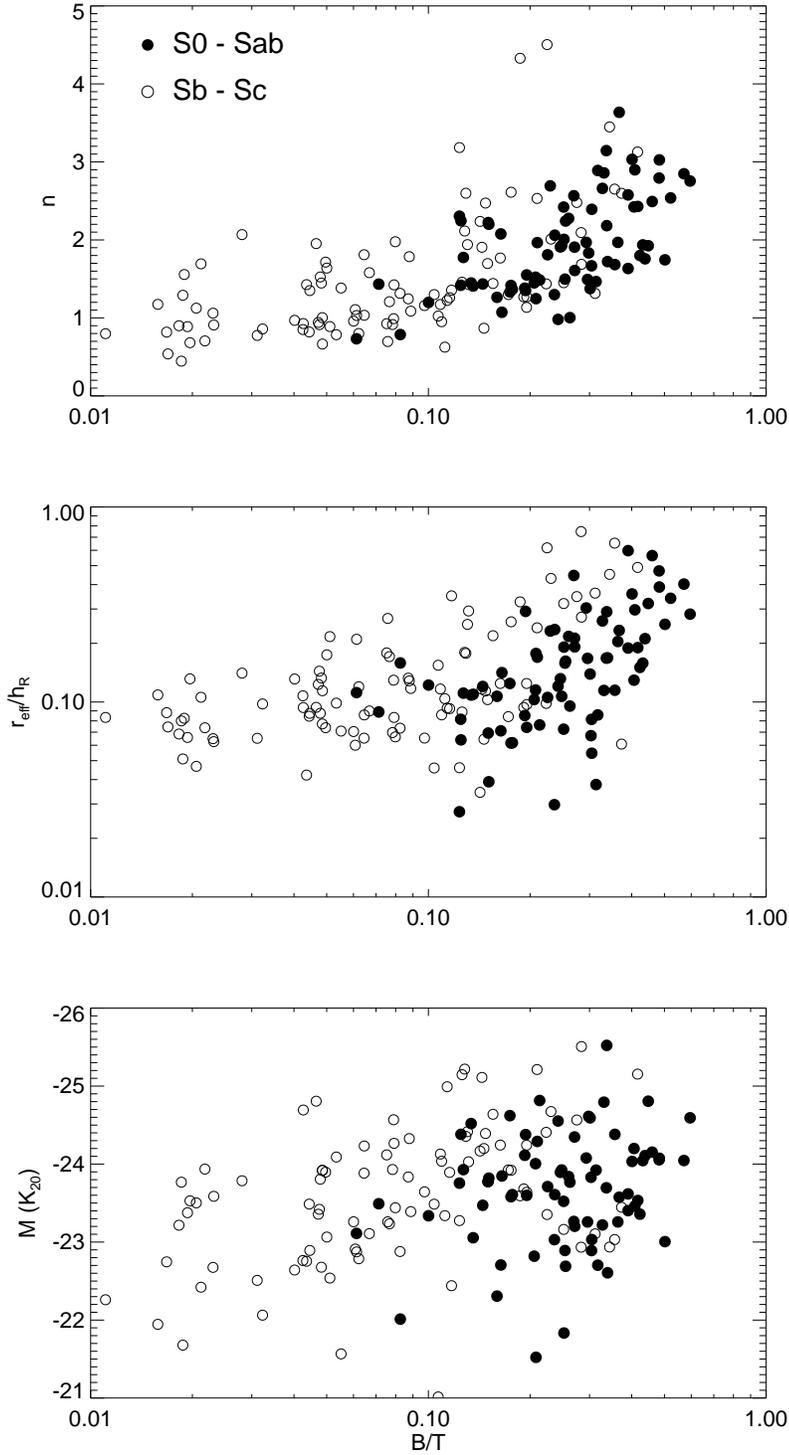}
\caption{For the same sample are shown the shape parameter of the bulge (upper panel),
the effective radius of the bulge, scaled to the scale length of the disk (middle panel), 
and the absolute galaxy luminosity (lower panel),
as a function of $B/T$. The scale lengths are measured from the near-IR images by applying
a multicomponent decompositions. 
The parameters are shown separately for the S0-Sab galaxies, and for the
Sb-Sc type spirals.  } 
\end{figure}

\clearpage

\begin{figure}
\includegraphics[]{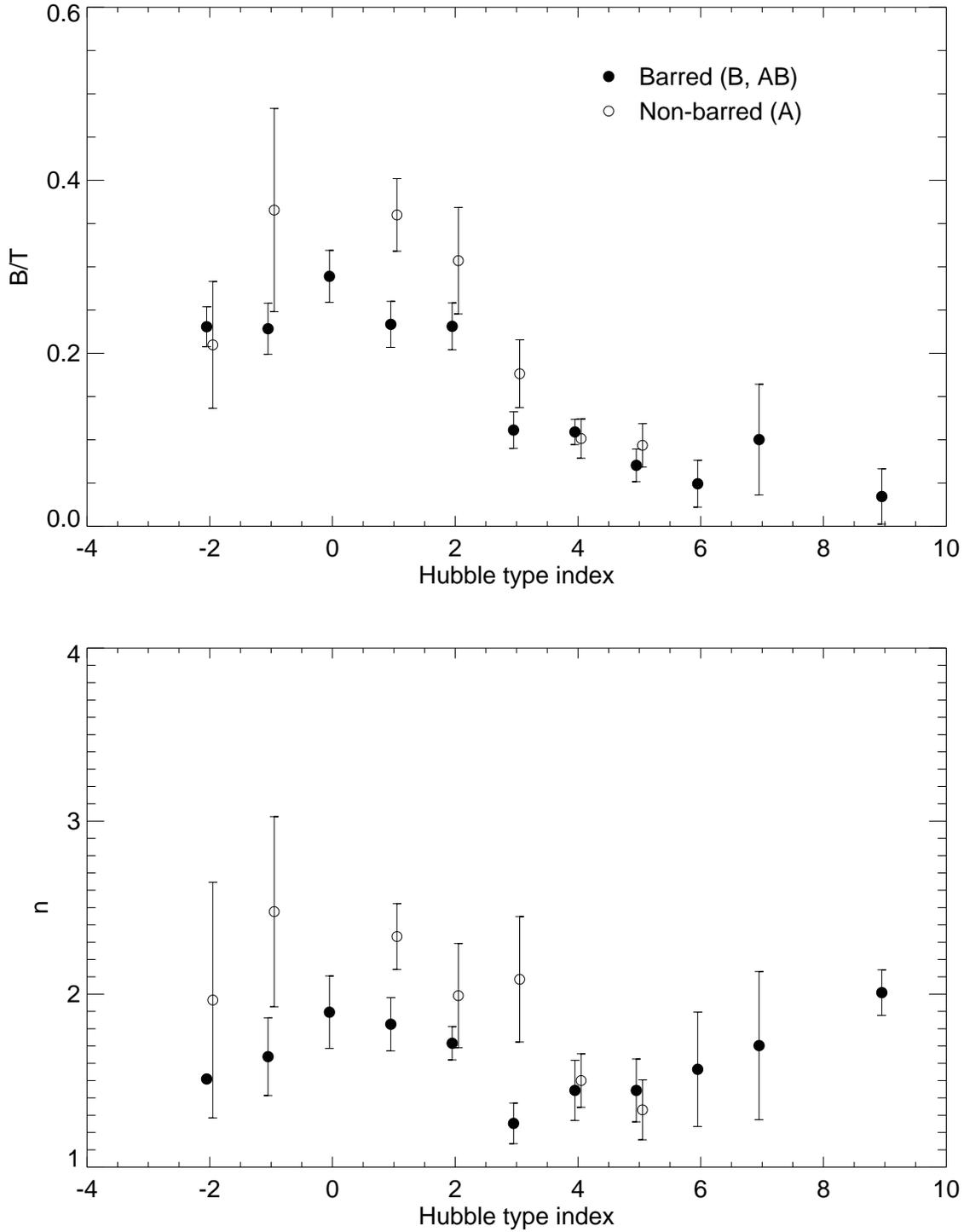}
\caption{ The parameters of the bulge, $B/T$ and $n$, are shown separately 
for barred and non-barred galaxies as a function of the Hubble type index. 
The classification of barred/non-barred is taken from ``The de Vaucouleurs Atlas of Galaxies''
by Buta et al. (2007). Again, the symbols are mean values and the error bars indicate the
standard deviations of the mean values.}
\end{figure}

\newpage
\clearpage
\begin{figure}
\includegraphics[]{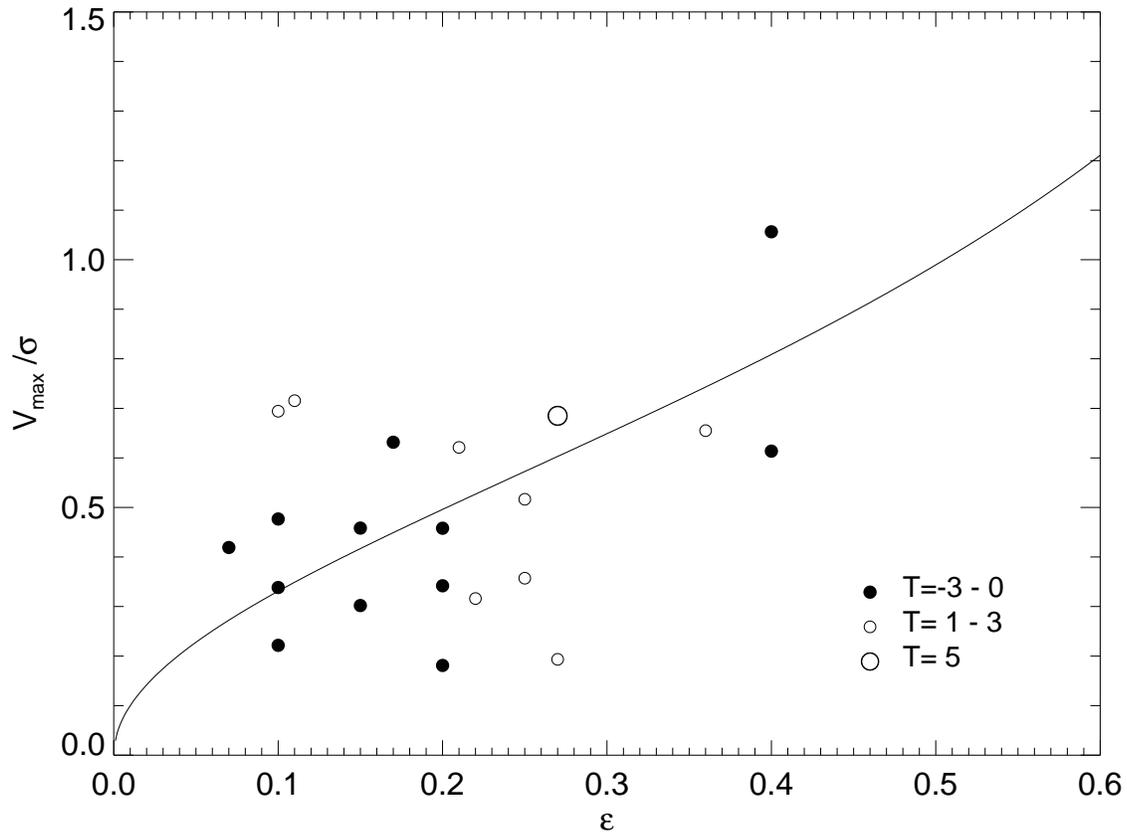}
\caption{ The kinematical properties of the bulges for a subsample of galaxies.
$V_{max}$ is the maximum line of sight rotation velocity of the bulge measured from the absorption lines,
$\sigma$ is the line of sight stellar velocity dispersion of the bulge just outside the nucleus, and
$\epsilon$ is the characteristic ellipticity in the region interior to the radius
of $V_{max}$. The values are taken from our Table 1, and the different symbols represent
different Hubble type indexes. Inclination of the disk would shift the data points
almost along the oblate rotator line, as predicted by the models by Binney (2005).
} 
\end{figure}

\clearpage
\begin{figure}
\includegraphics[width=84mm]{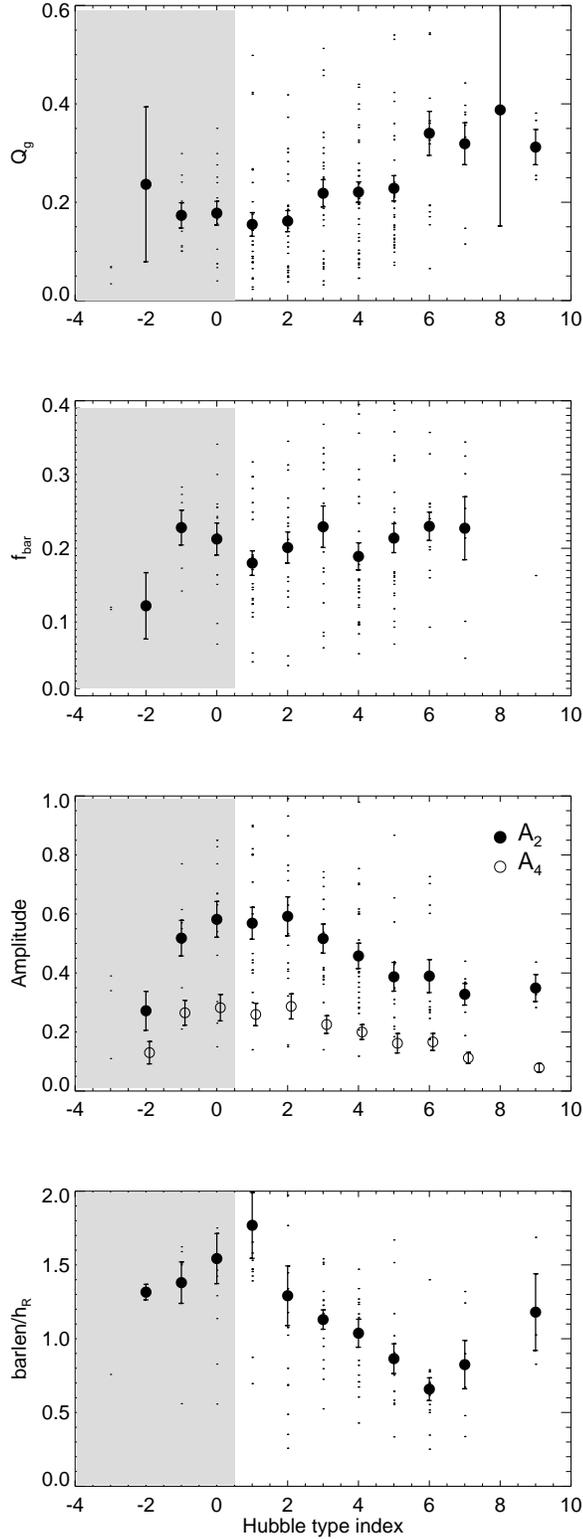}
\caption{ Four different estimates of bar strength are calculated for our sample 
of 216 galaxies and shown as a function of the Hubble type index. 
$Q_g$ is a bar torque indicator, which is the bar induced maximum tangential 
force, divided by the azimuthally averaged radial axisymmetric force field. The parameter
$f_{bar}$ is a measure of the ellipticity of the bar, as defined by Whyte et al. (2002),
and explained in more detail in the text. The $A_2$ and $A_4$ are amplitudes 
of density for the $m$=2 and $m$=4 Fourier modes. Bar length has been estimated
from the phases of the $A_2$ Fourier amplitudes so that the length is the radial
distance where the phase is maintained nearly constant in the bar region. The length is normalized
to the scale length of the disk, $h_R$, obtained from deep near-IR image. }
\end{figure}

\begin{figure}
\includegraphics[width=99mm]{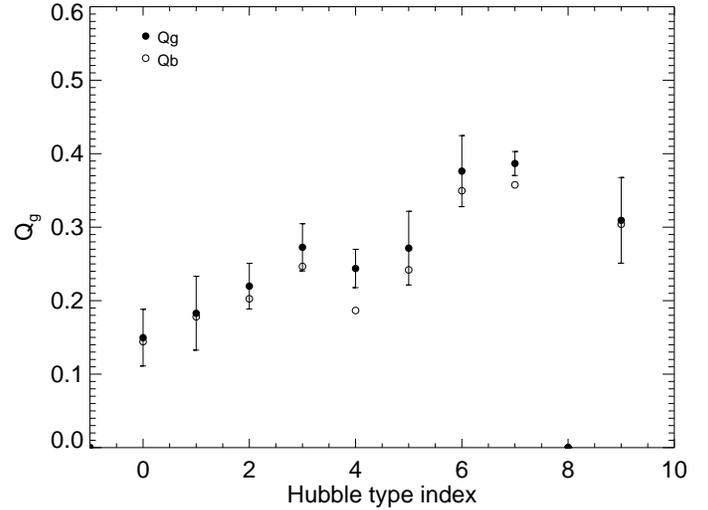}
\caption{Comparison of the mean bar torques with ($Q_b$, from Buta, Laurikainen $\&$ Salo
2004) and without ($Q_g$, from Buta et al. 2005) the correction
of the spiral arms in the OSUBGS sample. Both values are calculated adding the Fourier modes 
up to m=10. Only those galaxies are included for which it was possible to apply 
the bar/spiral separation approach.  
}
\end{figure}
 
\clearpage

\begin{figure}
\includegraphics[]{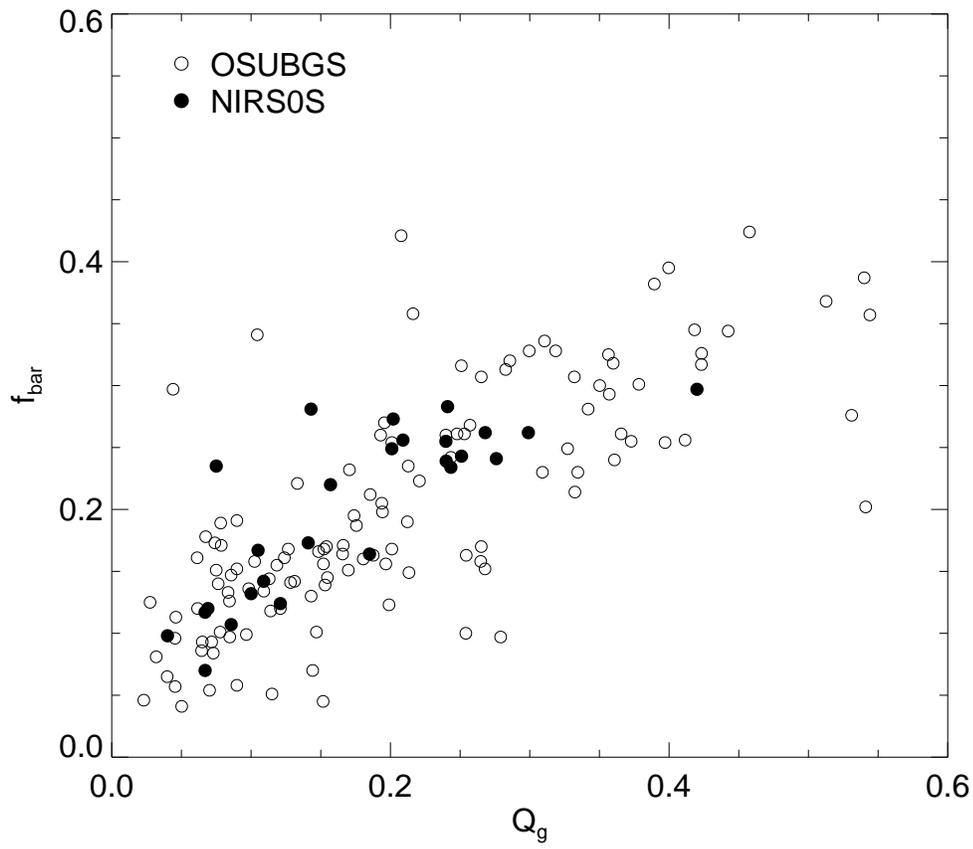}
\caption{ A correlation between the ellipticity of the bar, $f_{bar}$ and
the bar torque, $Q_g$. The open circles show the galaxies in the OSUBGS sample, 
whereas the filled circles show the galaxies in the NIRS0S sample.}
\end{figure}
\clearpage

\begin{figure}
\includegraphics[]{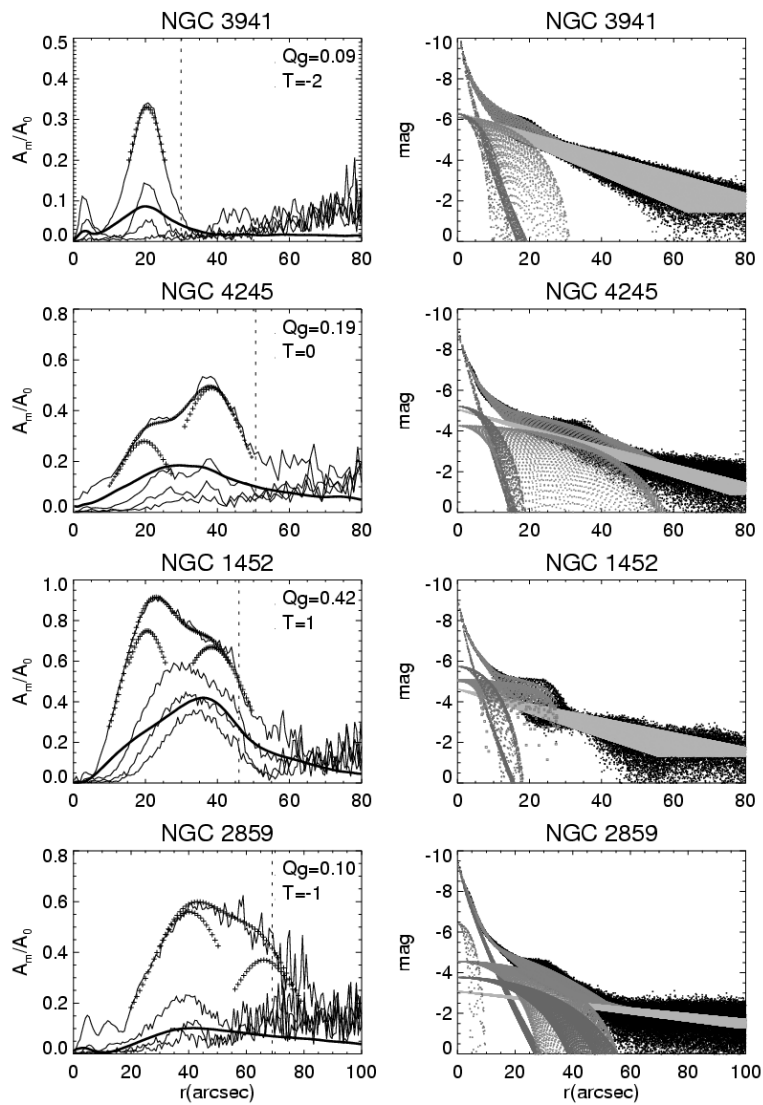}
\caption{ Examples of early-type disk galaxies in our sample. In the
left row are shown the Fourier amplitude profiles (thin lines) and
the $Q_T$ profiles (thick lines) for these galaxies. The crosses show
simple double Gaussian fits to the $A_2$ profiles, for which fits the parameters were
taken from the Table XX in Buta et al. (2006): shown separately
are the two peaks and the overall fit to the observed profile.
The vertical dashed line shows the length of the bar, estimated from
the phases of the $A_2$ Fourier amplitudes. 
In the upper right corners of these figures
are indicated the values of the bar torques and the Hubble type indexes.
In the y-axis the densities $A_m$ of each mode are divided by the axisymmetric,
azimuthally  averaged densities $A_0$. In the right row of the figure are shown the results
of the multicomponent decompositions for the surface brightness
profiles for the same galaxies. The disk was fitted by an exponential function,
the bulge by a Sersic's function, and the bars typically by a 
Ferrers function and the ovals and inner components of bars by a 
Sersic's function. A more detailed description of these decompositions can
be found in Laurikainen et al. (2006).}
\end{figure}
\clearpage

\begin{figure}
\includegraphics[]{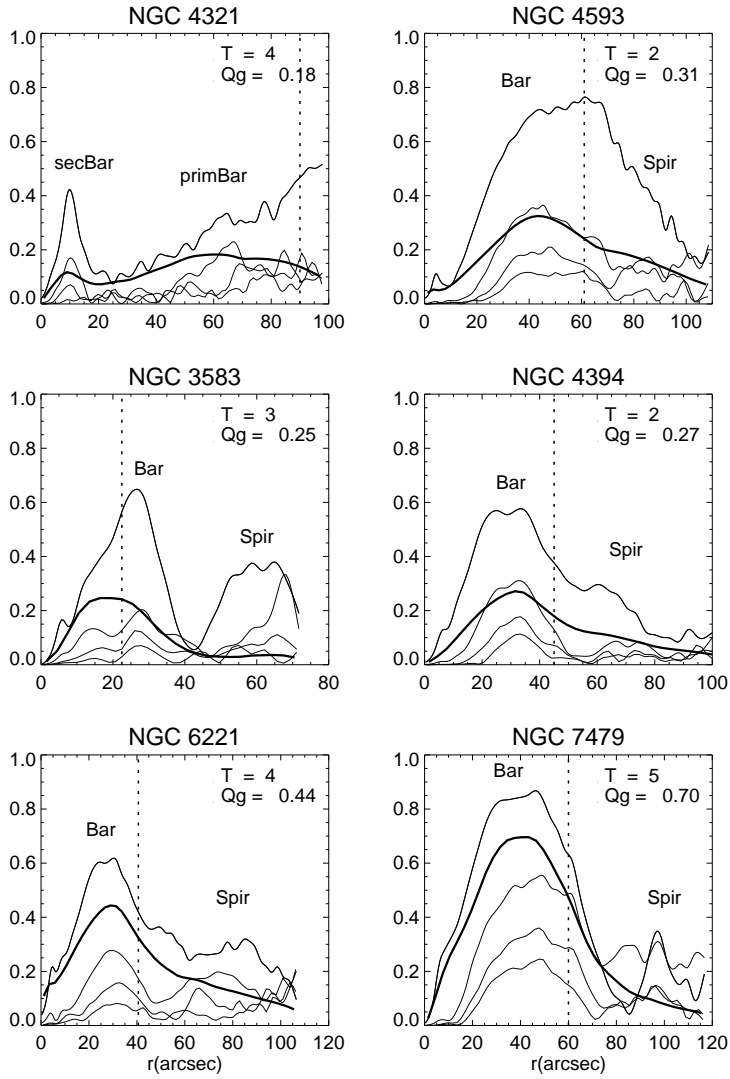}
\caption{ Characteristic examples of spiral galaxies in our sample. These figures
are similar to those shown in the left row in Figure 7. The symbols are also
the same as in Figure 7.}
\end{figure}


\begin{thebibliography}{99}

 \bibitem[\protect\citeauthoryear{Abraham et al.}{1999}]{abraham} Abraham, R. G., Merrifield, M. R., Ellis, R. S., Tanvir, N. R., Brinchmann, J. 1999, MNRAS, 308, 569
 \bibitem[\protect\citeauthoryear{Aguerri et al.}{2003}]{aquerri} Aguerri, J.A.L, Debattista, V.P., Corsini, E.M.  2003, MNRAS, 338, 465
 \bibitem[\protect\citeauthoryear{Andredakis \& Sanders}{1994}]{andredakis1994} Andredakis, Y.C., Sanders, R.H. 1994, MNRAS, 267, 283
 \bibitem[\protect\citeauthoryear{Andredakis, Peletier \& Balcells}{1995}]{andredakis1995} Andredakis, Y.C., Peletier, R.F., Balcells, M. 1995, MNRAS, 275, 874
 \bibitem[\protect\citeauthoryear{Athanassoula}{1992}]{atha1992} Athanassoula, E. 1992, MNRAS, 259, 345
 \bibitem[\protect\citeauthoryear{Athanassoula \& Beaton}{2006}]{atha2006b} Athanassoula, E., Beaton, R.L. 2006, MNRAS, 370, 1499
 \bibitem[\protect\citeauthoryear{Athanassoula \& Misioritis}{2002}]{atha2002} Athanassoula, E., Misioritis, A. 2002, MNRAS, 330, 35
 \bibitem[\protect\citeauthoryear{Athanassoula}{2003}]{atha2003} Athanassoula, E. 2003, MNRAS, 341, 1179 
 \bibitem[\protect\citeauthoryear{Athanassoula, Lambert \& Dehnen}{2005}]{atha2005a} Athanassoula, E., Lambert, J.C., Dehnen, W. 2005, MNRAS, 363, 496 
 \bibitem[\protect\citeauthoryear{Athanassoula}{2005}]{atha2005b} Athanassoula. E., 2005, MNRAS, 358, 1477 
 \bibitem[\protect\citeauthoryear{Athanassoula}{2006}]{atha2006} Athanassoula, E. 2006, astro-ph/10113
 \bibitem[\protect\citeauthoryear{Balcells \& Gonzalez}{1998}]{balcells1998} Balcells, M., Gonzalez, A.C. 1998, ApJ, 505, L109
 \bibitem[\protect\citeauthoryear{Balcells et al.}{2003}]{balcells2003} Balcells, M., Graham, A.W., Dominiguez-Palmero, L., Peletier, R.F. 2003, ApJ, 582, L79
 \bibitem[\protect\citeauthoryear{Balcells, Graham \& Peletier}{2004}]{balcells2004} Balcells, M., Graham, A.W., Peletier, R.F. 2004, astro-ph/0404379
 \bibitem[\protect\citeauthoryear{Barnes}{1988}]{barnes} Barnes, J.E. 1988, ApJ, 331, 699
 \bibitem[\protect\citeauthoryear{Batcheldor et al.}{2005}]{bat} Batcheldor, D., Axon, D., Merritt, D., Hughes, M. A., Marconi, A., Binney, J., 
                                    Capetti, A., Merrifield, M., Scarlata, C., Sparks, W. 2005, ApJS, 160, 76
 \bibitem[\protect\citeauthoryear{Bell \& de Jong}{2001}]{bell} Bell, E. F., de Jong, R.S. 2001, ApJ, 550, 212
 \bibitem[\protect\citeauthoryear{Benson, Frenk \& Sharples}{2002}]{benson} Benson, A.J., Frenk, C.S., Sharples, R.M. 2002, ApJ, 574, 104
 \bibitem[\protect\citeauthoryear{Berentzen et al.}{1998}]{berenzen} Berentzen, I., Heller, C.H., Shlosman, I., Fricke, K.J. 1998, MNRAS, 300, 49
 \bibitem[\protect\citeauthoryear{Bertin et al.}{1994}]{bertin} Bertin, G., Bertola, F., Buson, L. M., Danzinger, I. J., Dejonghe, H., Sadler, E. M., Saglia, R. P.,
        de Zeeuw, P. T., Zeilinger, W. W. 1994, AA, 292, 381
 \bibitem[\protect\citeauthoryear{Bettoni, Galletta \& Vallenari}{1988}]{bettoni1988} Bettoni, D., Galletta, G., Vallenari, A. 1988, AA, 197, 69
 \bibitem[\protect\citeauthoryear{Bettoni \& Galletta}{1997}]{bettoni1997} Bettoni, D., Galletta, G. 1997, AAS, 124, 61
 \bibitem[\protect\citeauthoryear{Binney}{1978}]{binney1978} Binney, J. 1978, MNRAS, 183, 501
 \bibitem[\protect\citeauthoryear{Binney}{2005}]{binney2005} Binney, J. 2005, astro-ph/0504387
 \bibitem[\protect\citeauthoryear{Binney \& Tremaine}{1987}]{binney1987} Binney, J., Tremaine, S. 1987, in ``Galactic Dynamics'' (Princeton: Princeton
       University Press)
 \bibitem[\protect\citeauthoryear{Bournaud \& Combes}{2002}]{bournaud} Bournaud, F., Combes, F. 2002, AA, 392, 83
 \bibitem[\protect\citeauthoryear{Bournaud, Combes \& Semelin}{2005}]{bournaud2005} Bournaud, F., Combes, F., Semelin, B. 2005, MNRAS, 364, L18 
 \bibitem[\protect\citeauthoryear{Bosma}{2003}]{bosma} Bosma, A. 2003, IAU Symp 220, 
 \bibitem[\protect\citeauthoryear{Bureau \& Athanassoula}{2005}]{bureau2005} Bureau, M., Athanassoula, E. 2005, ApJ, 626, 159
 \bibitem[\protect\citeauthoryear{Bureau et al.}{2006}]{bureau2006} Bureau, M., Aronica, G., Athanassoula, E., Dettmar, R.J., Bosma, A., Freeman, K.C. 2006, MNRAS, 370, 753

 \bibitem[\protect\citeauthoryear{Buta, Laurikainen \& Salo}{2004}]{buta2004} Buta, R., Laurikainen, E., Salo, H. 2004, AJ, 127, 279
 \bibitem[\protect\citeauthoryear{Buta et al.}{2005}]{buta2005} Buta, R, Vasylyev, S., Salo, H., Laurikainen, E. 2005, AJ, 130, 506
 \bibitem[\protect\citeauthoryear{Buta et al.}{2006}]{buta2006} Buta, R., Laurikainen, E., Salo, H., Block, D., Knapen, J. 2006, AJ, 132, 1859
 \bibitem[\protect\citeauthoryear{Buta et al.}{2007}]{buta2007} Buta, R., Corwin, H., Odewahn, S. 2007, ``The de Vaucouleurs Atlas of Galaxies'', Cambridge, 
                                         Cambridge University Press
 \bibitem[\protect\citeauthoryear{Caon, Cappaccioli \& D'Onofrio}{1993}]{caon} Caon, N., Cappaccioli, M., D'Onofrio, M. 1993, MNRAS, 265, 1013
 \bibitem[\protect\citeauthoryear{Carollo et al.}{1997}]{caroolo1997} Carollo, C.M., Stiavelli, M., de Zeeuw, P.T., Mack, J. 1997, AJ, 114, 2366
 \bibitem[\protect\citeauthoryear{Carollo, Stiavelli  \& Mack}{1998}]{carollo1998} Carollo, C., Stiavelli, M., Mack, J. 1998, AJ, 116, 68
 \bibitem[\protect\citeauthoryear{Carollo et al.}{2002}]{carollo2002} Carollo, C.M., Stiavelli, M., Seigar, M., de Zeeuw, P.T., Dejonghe, H. 2002, AJ, 123, 159
 \bibitem[\protect\citeauthoryear{Chung \& Bureau}{2004}]{chung} Chung, A., Bureau, M. 2004, ApJ, 127, 3192
 \bibitem[\protect\citeauthoryear{Combes et al.}{1990}]{combes1990} Combes, F., Debbash, F., Friedli, D., Pfenniger, D. 1990, AA, 233, 82
 \bibitem[\protect\citeauthoryear{Cappellari et al.}{2005}]{capplellari2005} Cappellari, M., Bacon, R., Bureau, M., Davies, R.L., de Zeeuw, P.T., Emsellem, E.,
                                     Falcon-Barroso, J., Krajnvic, D.,
                                     Kuntschner, H., McDermid, R.M., Peletier, R.F., Sarzi, M., van den Bosch, R.C.E., van den Ven, G. 2005,  
                                     Nearly Normal Galaxies in a LCDM Universe. A conference celebrating the 60th birthdays of George Blumenthal, 
                                     Sandra Faber and Joel
                                     Primack. 2005. Santa Cruz: UC Santa Cruz
\bibitem[\protect\citeauthoryear{Corsini et al.}{1999}]{corsini1999} Corsini, E. M., Pizzella, A., Sarzi, M., Cinzano, P., Vega Beltrán, J. C., Funes, J. G.,
          Bertola, F., Persic, M., Salucci, P. 1999, AA, 342, 671
 \bibitem[\protect\citeauthoryear{Corsini, Pizzella \&  Bertola}{2002}]{corsini2002} Corsini. E. M., Pizzella, A., Bertola, F. 2002, AA, 382, 488
 \bibitem[\protect\citeauthoryear{Corsini, Debattista \& Aguerri}{2003}]{corsini2003} Corsini,E. M., Debattista, V.P., Aguerri, J.A.L. 2003, ApJ, 599, L29
 \bibitem[\protect\citeauthoryear{Cowie et al.}{1996}]{cowie} Cowie, L. L., Songaila, A., Hu, E.M. 1996, AJ, 112, 839
 \bibitem[\protect\citeauthoryear{Davies et al.}{1983}]{davies1983} Davies, R.L., Efstathiou, G., Fall, S.M., Illingworth, G., Schechter, P.L. 1983, ApJ, 266, 41 
\bibitem[\protect\citeauthoryear{Debattista \& Williams}{2004}]{debattista2004} Debattista, V.P., Williams, T.B. 2004, AJ, 605, 714
 \bibitem[\protect\citeauthoryear{Debattista et al.}{2006}]{debattista2006} Debattista, V.P., Mayer, L., Carollo, C.M., Moore, B., Wadsley, J., 
                                      Quinn, T. 2006, ApJ, 645, 209
 \bibitem[\protect\citeauthoryear{de Grijs}{1998}]{degrijs} de Grijs, R. 1998, MNRAS, 299, 595
 \bibitem[\protect\citeauthoryear{Denicolo et al.}{2005}]{denicolo} Denicoló, G., Terlevich, R., Terlevich, E., Forbes, D. A., Terlevich, A., 
                                    Carrasco, L. 2005, MNRAS, 356, 1440
 \bibitem[\protect\citeauthoryear{de Souza, Gadotti \& dos Anjos}{2004}]{souza} de Souza, R. E.; Gadotti, D. A.; dos Anjos, S. 2004, ApJS, 153, 411 
 \bibitem[\protect\citeauthoryear{de Vaucouleurs et al.}{1991}]{devauc} de Vaucouleurs, G., de Vaucouleurs, A., Corwin, H.G., Jr., Buta, R., Paturel, G., Fouque, P. 
                                    1991, Third Reference Catalogue of Bright Galaxies (New York: Springer)(RC3)
 \bibitem[\protect\citeauthoryear{Eggen, Lynden-Bell \& Sandage}{1962}]{eggen} Eggen, O.J., Lynden-Bell, D. Sandage, A.R. 1962, ApJ, 136, 748
 \bibitem[\protect\citeauthoryear{Elmegreen \& Elmegreen}{1985}]{elm1985} Elmegreen, B., Elmegreen, D. 1985, ApJ, 288, 438
 \bibitem[\protect\citeauthoryear{Erwin}{2005}]{erwin} Erwin, P. 2005, MNRAS, 364, 283
 \bibitem[\protect\citeauthoryear{Eskridge et al.}{2002}]{eskridge} Eskridge, P.B. et al. 2002, ApJS, 142, 73
 \bibitem[\protect\citeauthoryear{Falcón-Barroso et al.}{2003}]{falcon2003} Falcon-Barroso, J., Balcells, M., Peletier, R.F., Vazdekis, A. 2003, AA, 405, 455
 \bibitem[\protect\citeauthoryear{Falcón-Barroso et al.}{2004}]{falcon2004} Falcón-Barroso, J. Peletier, R. F., Emsellem, E. Kuntschner, H., Fathi, K., Bureau, M.,
                Bacon, R., Cappellari, M., Copin, Y., Davies, R. L., de Zeeuw, T. 2004, MNRAS, 350, 35
 \bibitem[\protect\citeauthoryear{Fillmore, Boroson \& Dressler}{1986}]{fillmore} Fillmore, J.A., Boroson, T.A., Dressler, A. 1986, ApJ, 302, 208
 \bibitem[\protect\citeauthoryear{Fisher}{1997}]{fisher1997} Fisher, D. 1997, AJ, 113, 950
 \bibitem[\protect\citeauthoryear{Fisher}{2006}]{fisher2006} Fisher, D.B., 2006, ApJL, 642, 17
\bibitem[\protect\citeauthoryear{Fritze \& Alvensleben}{2004}]{alven2004} Fritze , U., Alvensleben, V. 2004, in ``Penetrating Bars through Masks of Cosmic
                                   Dust: The Hubble Turning Fork strikes a New Note'', ed. D.L. Block, I. Puerari, K.H. Freeman, R. Groess, E. Block 2004,
                                         Springer, p. 81 
 \bibitem[\protect\citeauthoryear{Gardner et al.}{1997}]{gardner2} Gardner, J.P., Sharples, R.M., Frenk, C.S., Carrasco, B.E. 1997, ApJ, 480, L99
 \bibitem[\protect\citeauthoryear{Gerssen et al.}{2003}]{gerssen} Gerssen, J., Kuijken, K., Merrifield, M.R. 2003, MNRAS, 345, 261
 \bibitem[\protect\citeauthoryear{Graham}{2001}]{graham} Graham, A.W. 2001, AJ, 121, 820
 \bibitem[\protect\citeauthoryear{Haynes et al.}{2000}]{haynes} Haynes, M.P., Jore, K.P., Barrett, E.A., Broeils, A.H., Murray, B.M. 2000, AJ, 120, 703
 \bibitem[\protect\citeauthoryear{Heraudeau \& Simien}{1998}]{heraudeau1998} Heraudeau, Ph., Simien, F. 1998, AAS, 133, 317
 \bibitem[\protect\citeauthoryear{Heraudeau et al.}{1999}]{heraudeau1999} Heraudeau, Ph., Simien, F., Maubon, G., Prugniel, Ph. 1999, AAS, 136, 509
 \bibitem[\protect\citeauthoryear{Hernquist}{1993}]{hern} Hernquist, L.E., 1993, ApJ, 409, 548
 \bibitem[\protect\citeauthoryear{Hozumi \& Hernquist}{2005}]{hozumi2005} Hozumi, S., Hernquist, L. 2005, PASP, 57, 719
 \bibitem[\protect\citeauthoryear{Illingworth}{1977}]{illing} Illingworth, G. 1977, ApJ, 218, L43
 \bibitem[\protect\citeauthoryear{Illingworth}{1981}]{illing1981} Illingworth, G. 1981, in ``Structure and Evolution of Normal Galaxies, ed. S.M. 
                                   Fall and D. Lynden-Bell (Cambridge: Cambridge University Press), p. 27
 \bibitem[\protect\citeauthoryear{Jarvis et al.}{1988}]{jarvis} Jarvis B.J., Dubath, P., Martinet, L., Bacon, R. 1988, AAS, 74, 513
 \bibitem[\protect\citeauthoryear{Jore, Broeils \& Haynes}{1996}]{jore} Jore, K.P., Broeils, A.H., Haynes, M.P. 1996, AJ, 112, 438
 \bibitem[\protect\citeauthoryear{Knapen et al.}{1995}]{1995} Knapen, J.~H., Beckman, J.~E., Heller, C.~H., Shlosman, I., $\&$ de Jong, R.~S. 1995, apj, 454, 623
 \bibitem[\protect\citeauthoryear{Kormendy}{1979}] {kormendy1979} Kormendy, J. 1979, ApJ, 227, 714
 \bibitem[\protect\citeauthoryear{Kormendy}{1981}]{kormendy1981} Kormendy, J. 1981, in ``Structure and Evolution of Normal Galaxies, ed. S.M. 
                                   Fall and D. Lynden-Bell (Cambridge: Cambridge University Press), p. 85
 \bibitem[\protect\citeauthoryear{Kormendy}{1982}] {kormendy1982} Kormendy, J. 1982, ApJ, 257, 75
 \bibitem[\protect\citeauthoryear{Kormendy}{1983}] {kormendy1983} Kormendy, J. 1983, ApJ, 275, 529
 \bibitem[\protect\citeauthoryear{Kormendy}{1984}]{kormendy1984} Kormendy, J. 1984, ApJ, 286, 116
 \bibitem[\protect\citeauthoryear{Kormendy}{1993}]{kormendy1993} Kormendy, J. 1993, in IAU Symp. 153: Galactic Bulges Vol 153, Kinematics of 
             extragalactic bulges: evidence that some galaxies are really disks, p. 209
 \bibitem[\protect\citeauthoryear{Kormendy \& Kennicutt}{2004}]{kormendy2004} Kormendy, J., Kennicutt, R.C. Jr. 2004, Ann Rev Astr Ap, Vol 42,603
 \bibitem[\protect\citeauthoryear{Kormendy \& Fisher}{2005}]{kormendy2005} Kormendy, J., Fisher, D. 2005, Rev MexAA, 23, 101
 \bibitem[\protect\citeauthoryear{Kormendy et al.}{2006}]{kormendy2006} Kormendy, J., Cornell, M.E., Block, D., Knapen, J.H., Allard, E.L. 2006, ApJ, 642, 765
 \bibitem[\protect\citeauthoryear{Kuijken \& Merrifield}{1995}]{kuijken} Kuijken, K., Merrifield, M.R., 1995, ApJ, 443, L13
 \bibitem[\protect\citeauthoryear{Laurikainen \& Salo}{2002}]{lauri2002} Laurikainen, E., Salo, H. 2002, MNRAS, 337, 1118
 \bibitem[\protect\citeauthoryear{Laurikainen, Salo \& Rautiainen}{2002}]{lauri2002b} Laurikainen, E., Salo, H., Rautiainen, P. 2002, MNRAS, 337, 880
 \bibitem[\protect\citeauthoryear{Laurikainen, Salo \& Buta}{2004}]{lauri2004a} Laurikainen, E., Salo, H., Buta, R. 2004, ApJ, 607,103
 \bibitem[\protect\citeauthoryear{Laurikainen et al.}{2004}]{lauri2004b} Laurikainen, E., Salo, H., Buta, R., Vasylyev, S. 2004, MNRAS, 355, 1251
 \bibitem[\protect\citeauthoryear{Laurikainen, Salo \& Buta}{2005}]{lauri2005} Laurikainen, E., Salo, H., Buta, R. 2005, MNRAS, 362, 1319
 \bibitem[\protect\citeauthoryear{Laurikainen et al.}{2006a}]{lauri2006a} Laurikainen, E., Salo, H., Buta, R., Knapen, J., Speltincx, T., Block, D. 2006, AJ, 132, 2634
 \bibitem[\protect\citeauthoryear{Laurikainen, Salo \& Buta}{2006}]{lauri2006b} Laurikainen, E., Salo, H., Buta, R. 2006, IAU Symp. 235, 9
 \bibitem[\protect\citeauthoryear{Lima-Neto \& Combes}{1995}]{lime} Lima-Neto, G.B., Combes, F. 1995, AA, 294, 657
 \bibitem[\protect\citeauthoryear{L\"utticke, Dettmar \& Pohlen}{2000}]{luticke} L\"uticke, R., Dettmar, R.J., Pohlen M. 2000, AAS, 145, 435
 \bibitem[\protect\citeauthoryear{Lynden-Bell \& Wood}{1968}]{lynden1968} Lynden-Bell, D., Wood, R. 1968, MNRAS, 138, 495
 \bibitem[\protect\citeauthoryear{Lynden-Bell \& Kalnajs}{1972}]{lynden1972} Lynden-Bell, D., Kalnajs, A.J. 1972, MNRAS, 157, 1
 \bibitem[\protect\citeauthoryear{Lynden-Bell \& Pringle}{1974}]{lynden1974} Lynden-Bell, D, Pringle, J.E. 1974, MNRAS, 168, 603
 \bibitem[\protect\citeauthoryear{Marinova \& Jogee}{2006}]{marinova2006} Marinova, I., Jogee, S. 2006, astroph/0608039
 \bibitem[\protect\citeauthoryear{Martin}{1995}]{martin} Martin, P. 1995, AJ, 109, 2428
 \bibitem[\protect\citeauthoryear{Martinez-Valpuesta \& Shlosman}{2004}]{valpuesta2004} Martinez-Valpuesta, I., Shlosman, I. 2004, ApJL, 613, 29 
 \bibitem[\protect\citeauthoryear{Martinez-Valpuesta, Shlosman \& Heller}{2006}]{valpuesta2006} Martinez-Valpuesta, I., Shlosman, I., Heller, C. 2006, ApJ, 637, 214
 \bibitem[\protect\citeauthoryear{Mathieu, Merrifield \& Kuijken}{2002}]{mathieu} Mathieu, A., Merrifield, M.R.,Kuijken, K. 2002, MNRAS, 330, 251
 \bibitem[\protect\citeauthoryear{Magrelli, Bettoni \& Galletta}{1992}]{magrelli} Magrelli, G., Bettoni, D., Galletta, G. 1992, MNRAS, 256, 500
 \bibitem[\protect\citeauthoryear{Merrifield \& Kuijken}{1995}]{merrifield} Merrifield, M.R., Kuijken, K.1995, MNRAS, 274, 933
 \bibitem[\protect\citeauthoryear{Moiseev, Valdés \& Chavushyan}{2004}]{moiseev} Moiseev, A. V., Valdés, J. R., Chavushyan, V. H. 2004, AA, 421, 433
 \bibitem[\protect\citeauthoryear{Navarro \& Steinmetz}{2000}]{stenmtz} Navarro, J.F., Steinmetz, M. 2000, ApJ, 538, 477
 \bibitem[\protect\citeauthoryear{Ohta, Hamabe \& Wakamatsu}{1990}]{ohta1990} Ohta, K., Hamabe, M., Wakamatsu, K. 1990, ApJ, 357, 71
 \bibitem[\protect\citeauthoryear{Ohta}{1996}]{ohta1996} Ohta, K., in IAU Coll. 157 ``Barred Galaxies'', p. 37, eds. R. Buta, D.S. Crocker and B.G Elmegreen
 \bibitem[\protect\citeauthoryear{O'Neil \& Dubinski}{2003}]{oneil} O'Neil J.K., Dubinski J. 2003, MNRAS, 251
 \bibitem[\protect\citeauthoryear{Palacios et al.}{1997}]{palacios} Palacios, J., Garcia-Vargas, M. L., Diaz, A., Terlevich, R., Terlevich, E. 1997, AA, 323, 749
 \bibitem[\protect\citeauthoryear{Patsis, Skokos \& Athanassoula}{2002}]{patsis} Patsis, P.A., Skokos, Ch., Athanassoula, E., 2002, MNRAS, 337, 578
 \bibitem[\protect\citeauthoryear{Persic, Salucci \& Stel}{1996}]{persic} Persic, M., Salucci, P., Stel, F. 1996, MNRAS, 281, 27
 \bibitem[\protect\citeauthoryear{Pfenniger \& Friedli}{1991}]{pfenniger} Pfenniger, D., Friedli, D. 1991, AA, 252, 75 
 \bibitem[\protect\citeauthoryear{Pizzella et al.}{2004}]{pizzella} Pizzella, A., Corsini, E.M., Vega Beltrán, J.C., Bertola, F.  2004, AA, 424, 447
 \bibitem[\protect\citeauthoryear{Raha et al.}{1991}]{raha} Raha. N., Sellwood, J.A., James, R.A., Kahn, F.D. 1991, Nature, 352, 411
 \bibitem[\protect\citeauthoryear{Rand \& Wallin}{2004}]{rand} Rand, R.J., Wallin, J.F. 2004, ApJ, 614, 412
 \bibitem[\protect\citeauthoryear{Rautiainen, Salo \& Laurikainen}{2005}]{rauti} Rautiainen, P., Salo, H., Laurikainen, E., 2005, ApJ, 631, L129
 \bibitem[\protect\citeauthoryear{Regan \& Elmegreen}{1997}]{regan} Regan, M.W., Elmegreen, D.M. 1997, AJ, 114, 965
 \bibitem[\protect\citeauthoryear{Romanowsky}{2006}]{romanowsky} Romanowsky, A.J. 2006, IAU Symp, 234, ``Planetary Nebulae in Our Galaxy and Beyond'', eds. 
            M.J. Barlow, R.H. Mendez (Cambridge Univ. Press)
 \bibitem[\protect\citeauthoryear{Sakamoto et al.}{1999}]{sakamoto} Sakamoto, K., Okumura, S. K., Ishizuki, S. Scoville, N. Z. 1999, ApJ, 525, 691
 \bibitem[\protect\citeauthoryear{Salo et al.}{1999}]{salo1999} Salo, H., Rautiainen, P., Buta, R., Purcell, G., Cobb, M., Crocker, D. A., Laurikainen, E., 1999, 
AJ, 117, 792.
 \bibitem[\protect\citeauthoryear{Salo et al.}{2006}]{salo2006} Salo, H., Laurikainen, E., Rautiainen, P., Buta R. 2006, IAU Symp, 235, 347
 \bibitem[\protect\citeauthoryear{Schechter \& Dressler}{1987}]{Schechter} Schechter, P. L., Dressler, A. 1987, AJ, 94, 563
 \bibitem[\protect\citeauthoryear{Schultz et al.}{2003}]{schultz} Schultz, J., Fricke, U., Alvensleben, V., Fricke, K.J. 2003, AA, 389, 89
 \bibitem[\protect\citeauthoryear{Schweizer}{2005}]{schweizer} Schweizer, F. 2005, in ``Starbursts: From 30 Doradus to Lyman Break Galaxies, eds.
                                     de Grijs R.M., Gonzalez Delgado (Dordrecht:Springer), 143
 \bibitem[\protect\citeauthoryear{Sellwood}{1980}]{sellwood} Sellwood, J.A. 1980, AA, 89, 296
 \bibitem[\protect\citeauthoryear{Shapiro, Gerssen \& Marel}{2003}]{shapiro} Shapiro, K.L., Gerssen, J., van der Marel, R. P. 2003, AJ, 126, 2707
 \bibitem[\protect\citeauthoryear{Sheth et al.}{2005}]{sheth} Sheth, K., Vogel, S. N., Regan, M. W., Thornley, M. D., Teuben, P. J. 2005, ApJ, 632, 217
 \bibitem[\protect\citeauthoryear{Shlosman, Peletier \& Knapen}{2000}]{shlosman} Shlosman, I., Peletier, R.F., Knapen, J.H. 2000, ApJ, 535, L83
 \bibitem[\protect\citeauthoryear{Simien \& de Vaucouleurs}{1986}]{simien1986} Simien, F., de Vaucouleurs, G. 1986, ApJ, 302, 564
 \bibitem[\protect\citeauthoryear{Simien \& Prugniel}{1997}]{simien1997} Simien,F., Prugniel, Ph. 1997, AAS, 126, 15
 \bibitem[\protect\citeauthoryear{Springel \& Hernquist}{2005}]{springel} Springel, V., Hernquist, L. 2005, ApJ, 622, L9
 \bibitem[\protect\citeauthoryear{Steffen et al.}{2003}]{steffen2003} Steffen, A.T., Barger, A.J., Cowie, L.L., Mushotsky, R.F., Yaung, Y. 2003, ApJ, 596, L23
 \bibitem[\protect\citeauthoryear{Steinmetz \& Navarro}{2002}]{stein2002} Steinmetz, M., Navarro, J.F. 2002, NewA, 7, 155 
 \bibitem[\protect\citeauthoryear{Toomre}{1972}]{toomre1972} Toomre, A., Toomre, J. 1972, ApJ, 178, 623
\bibitem[\protect\citeauthoryear{Toomre}{1977}]{toomre1977} Toomre, A. 1977, sgsp.conf..401T
 \bibitem[\protect\citeauthoryear{Ueda et al.}                               {2003}]{ueda2003} Ueda, Y., Masayuki, A., Ohta, K., Miyaji, T. 2003, ApJ, 598, 886
 \bibitem[\protect\citeauthoryear{White \& Rees}{1978}]{white} White, S.D. M., Rees, M.J. 1978, MNRAS, 183, 341
 \bibitem[\protect\citeauthoryear{Whyte et al.}{2002}]{whyte} Whyte, L., Abraham, R.G., Merrifield, M.R., Eskridge, P.B., Frogel, J.A., Pogge, R.W. 2002, MNRAS, 336, 1281

\end{thebibliography}
\end{document}